\documentclass[twocolumn,preprintnumbers,nofootinbib,prd,superscriptaddress,groupedaddress,aps]{revtex4-1}

\usepackage{graphicx,amssymb,amsmath,amsthm,amsfonts,epstopdf,epsfig,epsf}
\usepackage[linktocpage]{hyperref}
\usepackage[usenames]{color}
\usepackage{epstopdf}

\usepackage{aas_macros}
\usepackage{bm}
\usepackage{dcolumn}
\usepackage[latin1]{inputenc}
\usepackage{latexsym}
\usepackage{rotating}
\usepackage{hyperref}
\usepackage{color}
\usepackage{longtable}
\usepackage{enumerate}
\usepackage{tensor}
\usepackage{stmaryrd}

\usepackage{url}
\def\pa{\partial}

\def\nn{\nonumber}

\newcommand{\ben}{\begin{enumerate}}
\newcommand{\een}{\end{enumerate}}

\def\be{\begin{equation}}
\def\ee{\end{equation}}
\newcommand{\beq}{\begin{eqnarray}}
\newcommand{\eeq}{\end{eqnarray}} 
\newcommand{\ba}{\begin{align}}
\newcommand{\ea}{\end{align}}

\def\l{\left}
\def\r{\right}

\def\nn{\nonumber}

\def\o{\omega}
\def\ba{\bar{a}}

\begin{document}

\title{Echoes of ECOs: gravitational-wave signatures of exotic compact objects and of quantum corrections at the horizon scale}

% Echoes of ECOs: Evidence for horizons and exotic compact objects in gravitational waveforms

\author{
Vitor Cardoso$^{1,2,3}$, %\footnote{Electronic address: vitor.cardoso@tecnico.ulisboa.pt},
Seth Hopper$^{1}$,
Caio F. B. Macedo$^{1}$,
Carlos Palenzuela$^{4}$,
Paolo Pani$^{5,1}$
}
\affiliation{${^1}$ CENTRA, Departamento de F\'{\i}sica, Instituto Superior T\'ecnico -- IST, Universidade de Lisboa -- UL,
Avenida Rovisco Pais 1, 1049 Lisboa, Portugal}
\affiliation{${^2}$ Perimeter Institute for Theoretical Physics, 31 Caroline Street North
Waterloo, Ontario N2L 2Y5, Canada}
\affiliation{${^3}$ Theoretical Physics Department, CERN, CH-1211 Geneva 23, Switzerland}
\affiliation{${^4}$Departament  de  F\'{\i}sica $\&$ IAC3,  Universitat  de  les  Illes  Balears  and  Institut  d'Estudis
Espacials  de  Catalunya,  Palma  de  Mallorca,  Baleares  E-07122,  Spain}
\affiliation{$^{5}$ Dipartimento di Fisica, ``Sapienza'' Universit\`a di Roma \& Sezione INFN Roma1, P.A. Moro 5, 00185, Roma, Italy}

\begin{abstract}
Gravitational waves from binary coalescences provide one of the cleanest signatures of the nature of compact objects. 
% If the objects are black holes, a smoking-gun evidence of their horizon is given by the high merger frequency and by the post-merger ringdown. 
It has been recently argued that the post-merger ringdown waveform of exotic ultracompact objects is initially identical to that of a black-hole, and that putative corrections at the horizon scale will appear as secondary pulses after the main burst of radiation. Here we extend this analysis in three important directions: (i)~we show that this result applies to a large class of exotic compact objects with a photon sphere for generic orbits in the test-particle limit; (ii)~we investigate the late-time ringdown in more detail, showing that it is universally characterized by a modulated and distorted train of ``echoes''of the modes of vibration associated with the photon sphere; (iii)~we study for the first time equal-mass, head-on collisions of two ultracompact boson stars and compare their gravitational-wave signal to that produced by a pair of black-holes. If the initial objects are compact enough as to mimic a binary black-hole collision up to the merger, the final object exceeds the maximum mass for boson stars and collapses to a black-hole. This suggests that --~in some configurations~-- the coalescence of compact boson stars might be almost indistinguishable from that of black-holes. On the other hand, generic configurations display peculiar signatures that can be searched for in gravitational-wave data as smoking guns of exotic compact objects.
\end{abstract}

%\tableofcontents
%\end{widetext}
%\clearpage

% \pacs{04.70.-s,04.80.-y,12.60.-i,11.10.St}
%14.80.-j 	Other particles (including hypothetical)
%11.10.St 	Bound and unstable states; Bethe-Salpeter equations
%12.60.-i 	Models beyond the standard model (for unified field theories, see 12.10.-g)
%04.25.D-    Numerical relativity
%04.25.dc    Numerical studies of critical behavior, singularities, and cosmic censorship
%04.25.dg    Numerical studies of black holes and black-hole binaries
%04.25.-g    general relativity: approximation methods, equations of motion
%04.50.-h    Higher-dimensional gravity and other theories of gravity
%04.50.Cd    KaluzaKlein theories
%04.50.Gh    Higher-dimensional black holes, black strings, and related objects
%04.60.Cf    Gravitational aspects of string theory
%04.70.-s    Physics of black holes
%04.70.Bw    Classical black holes
%04.70.Dy    Quantum aspects of black holes, evaporation, thermodynamics
%04.80.-y    Experimental studies of gravity
%04.80.Cc    Experimental tests of gravitational theories
%11.25.Mj    Compactification and four-dimensional models
%11.10.Kk    Field theories in dimensions other than four

\maketitle

% \tableofcontents
%%%%%%%%%%%%%%%%%%%%%%%%%%%%%%%%%%%%%%%%%%%%%%%
\section{Introduction}
%%%%%%%%%%%%%%%%%%%%%%%%%%%%%%%%%%%%%%%%%%%%%%%

The relativistic collision of two compact objects is the Rosetta Stone of the strong-gravity regime. The gravitational-wave (GW) signal emitted during the process contains a wealth of information on the nature of the colliding bodies. Following the recent LIGO detections~\cite{GW150914,GW151226,Abbott:2016bqf}, in the next years GW astronomy will deepen our understanding of the gravitational interaction and of astrophysics in extreme-gravity conditions to unprecedented level, playing a role similar to that of atomic spectroscopy in advancing quantum theory during the past century.

The comparison to atomic spectroscopy seems particularly apt at least in two respects: (i)~the post-merger ringdown phase is governed by a series of damped oscillatory modes~\cite{Vishveshwara:1970zz,Kokkotas:1999bd,Berti:2009kk} that can be computed very precisely in perturbation theory, and are akin to the energy levels of the hydrogen spectrum; (ii)~the precise modelling of the gravitational waveform allows us to search for smoking-gun anomalies due to new physics, similarly to the celebrated Lamb shift in atomic spectroscopy.

GW spectroscopy will play an increasingly important role as more and more events at large signal-to-noise ratio are detected. These observations provide novel ways to test strong gravity~\cite{Yunes:2013dva,Berti:2015itd,TheLIGOScientific:2016src,Yunes:2016jcc}, black-hole (BH) no-hair results~\cite{Cardoso:2016ryw}, the existence of event horizons~\cite{Cardoso:2016rao}, possible quantum effects at the horizon scale~\cite{Barausse:2014tra,Cardoso:2016rao}, dark matter and environmental effects~\cite{Barausse:2014tra,Macedo:2013qea}, and also exotic compact objects (ECOs) which might reveal themselves for the first time in the GW band~\cite{Macedo:2013jja,Cardoso:2016rao,Giudice:2016zpa}. All these tests require a precise modeling of the gravitational waveform in strong-gravity processes.

It has been recently argued~\cite{Cardoso:2016rao} that the post-merger ringdown phase of an ECO in the high-compactness limit is initially almost identical to that of a BH, and that any correction at the horizon scale due to a surface~\cite{Mazur:2004fk} or to quantum effects~\cite{Lunin:2001jy,Skenderis:2008qn,Almheiri:2012rt,Saravani:2012is} will reveal itself in secondary pulses that appear in the late-time ringdown waveform. This result was obtained by studying the radial plunge of a test particle into a thin-shell wormhole. If the wormhole throat is located at $r_0\sim 2M$, the initial ringdown signal is due to the vibration modes of the photon sphere (PS), 
and is the same as those of BHs, even though their quasinormal mode (QNM) spectrum - defined as the poles of the relevant Green's function~\cite{Kokkotas:1999bd,Berti:2009kk} - differ dramatically. BHs have QNMs which can be identified with the PS. There being no other scale in the problem and with ingoing conditions at the horizon, the PS modes are identical to the QNMs and no other mode is excited. For ECOs, on the other hand, the PS modes still exist and they ring in the same way as BHs, but are not QNMs, as they do not belong to the spectrum of the relevant operator anymore. Instead, the spectrum contains a series of trapped modes, which describe the vibration of the inner \textit{stable} PS, which is absent in BH spacetimes. 

With the exception of BSs - which we know to form naturally as a consequence of collapse of scalar fields - there is no known formation mechanisms for ECOs. In addition, there are some indications that horizonless, ultracompact objects (what we have termed as ECOs), are linearly or nonlinearly unstable~\cite{Cardoso:2014sna} although the timescales involved are model-dependent or unknown. 
Nevertheless, the time has come to expect the unexpected, and a good understanding of alternatives allows us to search for new physics in GW data. The understanding of ECOs is also important to quantify our confidence in the existence of BHs and event horizons.

The purpose of this work is to extend the analysis of Ref.~\cite{Cardoso:2016rao} in different independent directions. On the one hand, we consider different models of ECOs, generic trajectories of test particles, as well as the scattering of Gaussian wavepackets off these objects, showing that a generic feature of microscopic-scale corrections near the horizon is the presence of a modulated series of ``echoes''of the PS vibration modes. On the other hand we study, for the first time, 
the head-on collision of two solitonic boson stars (BSs) with a self-interacting potential including terms up to sixth order in the scalar field~\cite{Friedberg:1986tq,Lee:1991ax}. The latter are chosen because they can reach a compactness comparable to that of the PS, are relatively easy to evolve numerically, and can naturally form in dynamical scenarios~\cite{Liebling:2012fv}. To the best of our knowledge, no other model of ECO is known to enjoy all these properties. Thus, solitonic BSs stand out as the most natural model of ECOs and an important question is whether they can mimic the GW signal of a BH-BH coalescence. Through this work we use $c=G=1$ units.

%%%%%%%%%%%%%%%%%%%%%%%%%%%%%%%%%%%%%%%%%%%%%%%
\section{Echoes of ECOs}
%%%%%%%%%%%%%%%%%%%%%%%%%%%%%%%%%%%%%%%%%%%%%%%
In this section, we investigate the (ringdown) response of several models of ECOs in different scattering processes.
Most of our results are derived for the same wormhole model studied in Ref.~\cite{Cardoso:2016rao}, for thin-shell gravastars~\cite{Mazur:2004fk,Visser:2003ge}, 
%for Florides' fluid stars~\cite{Florides529}, 
and for a simple toy model of an empty, spherical thin shell of matter (model~II of matter-bumpy BH in Ref.~\cite{Barausse:2014tra} with $M=0$). 
The list is not meant to be exaustive, but merely to show that, qualitatively, the response of ultracompact objects is universal and simple, regardless of the specifics of the object.
All these models are characterized by some exotic form of matter that prevents gravitational collapse and, most importantly, have a radius $r_0$ that can be arbitrarily close to the would-be Schwarzschild radius. We focus on models in which 
\be
r_0=2M+\ell\,,
\ee
with $\ell\ll M$, which can qualitatively describe putative microscopic corrections at the horizon scale [cf., e.g., Refs.~\cite{Mazur:2004fk,Lunin:2001jy,Damour:2007ap,Skenderis:2008qn,Almheiri:2012rt} for some proposals].

In the spherically symmetric case, the line element for these models can be collectively written as
\begin{equation}
 ds^2=-F(r)dt^2+\frac{1}{B(r)}dr^2 +r^2d\Omega^2\,,\label{ds2}
\end{equation}
where $F$ and $B$ depend on the model~\cite{Cardoso:2014sna,Barausse:2014tra,Cardoso:2016rao}. Matter is localized only in the region $r\leq r_0$ whereas, in the region $r>r_0$, Birkhoff's theorem guarantees that spherically symmetric solutions are described by the Schwarzschild metric, $F(r)=B(r)=1-2M/r$. Details on each model are given in Appendix~\ref{app:models}. 
%%%%%%%%%%%%%%%%%%%%%%%%%%%%%%%%%%%%%%%%%%%%%%%
\subsection{Scattering of wavepackets\label{sec:ECO}}
\begin{figure}[th]
\begin{center}
\includegraphics[width=0.49\textwidth]{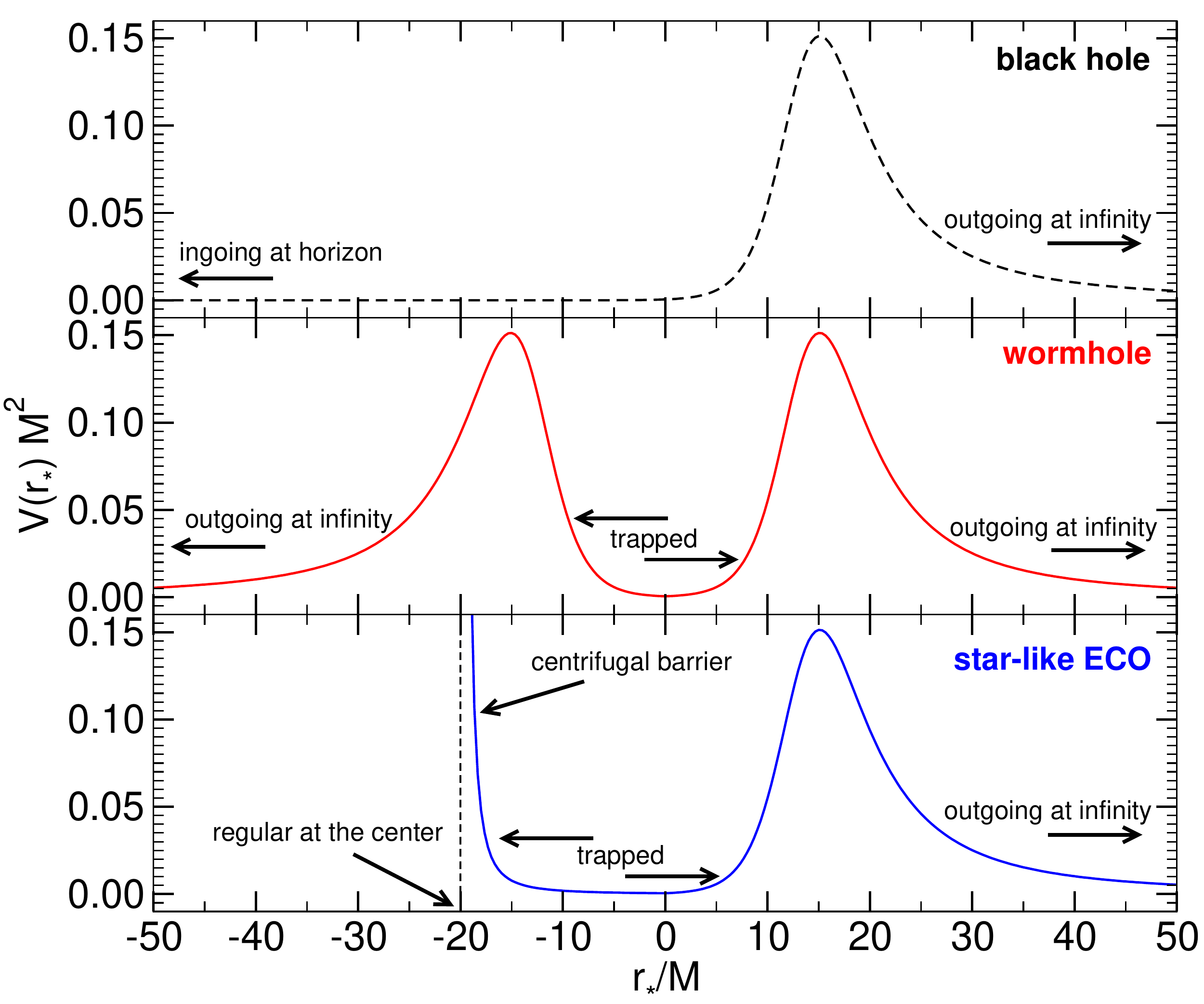}
\caption{Qualitative features of the effective potential felt by perturbations of a Schwarzschild BH compared to the case of wormholes~\cite{Cardoso:2016rao} and of star-like ECOs with a regular center~\cite{Cardoso:2014sna}. The precise location of the center of the star is model-dependent and was chosen for visual clarity.
The maximum and minimum of the potential corresponds approximately to the location of the unstable and stable PS,
and the correspondence is exact in the eikonal limit of large angular number $l$.
In the wormhole case, modes can be trapped between the PSs in the two ``universes''. In the star-like case, modes are trapped between the PS and the centrifugal barrier near the center of the star~\cite{1991RSPSA.434..449C,Chandrasekhar:1992ey,Abramowicz:1997qk}. In all cases the potential is of finite height, and the modes leak away, with higher-frequency modes leaking on shorter timescales. 
\label{fig:potential}}
\end{center}
\end{figure}

%%%%%%%%%%%%%%%%%%%%%%%%%%%%%%%%%%%%%%%%%%%%%%%
The most relevant signatures of these models are already evident in the simplest scattering process, namely a test scalar wavepacket being scattered off the gravitational potential of the ECO. The scattering is governed by the (Klein-Gordon) master equation~\cite{Berti:2009kk} $\Box \Phi=0$. Using angular variables $(\theta,\phi)$ on the sphere, and expanding the scalar in spherical harmonics as $\Phi=\sum_{lm}Y_{lm}(\theta,\phi)\Psi_{lm}(r)/r$ we get
%%%
\begin{align}
% \left[-\frac{\pa^2}{\pa t^2} + \frac{\pa^2}{\pa r_*^2} - V_\ell(r)\right]
% \Psi_{\ell m}(t,r) = 0\,,\label{eqn:TDmastereqn0}
\left[-\frac{\pa^2}{\pa t^2} + \frac{\pa^2}{\pa r_*^2} - V_l(r)\right]
\Psi_{l m}(t,r) = 0\,,\label{eqn:TDmastereqn}
\end{align}
%%%%
where $dr/dr_*=\sqrt{FB}$. In the exterior region, $r>r_0$, the tortoise coordinate $r_*$ and the potential $V_l$ coincide with their Schwarzschild values, 
% $r_*= r + 2M \log (r/2M - 1)$ and $V_l=(1-2M/r)(l(l+1)/r^2+2M/r^3)$, respectively, 
whereas their expressions in the interior region $r<r_0$ are model-dependent [cf. Appendix~\ref{app:models}].
%
% As an initial condition, we consider a Gaussian wavepacket $\Psi_{lm}\sim r^l$ near the center of the ECO.
%%%%
The potential for some representative cases is shown in Fig.~\ref{fig:potential}.

\begin{figure*}[th]
\begin{center}
\includegraphics[width=0.49\textwidth]{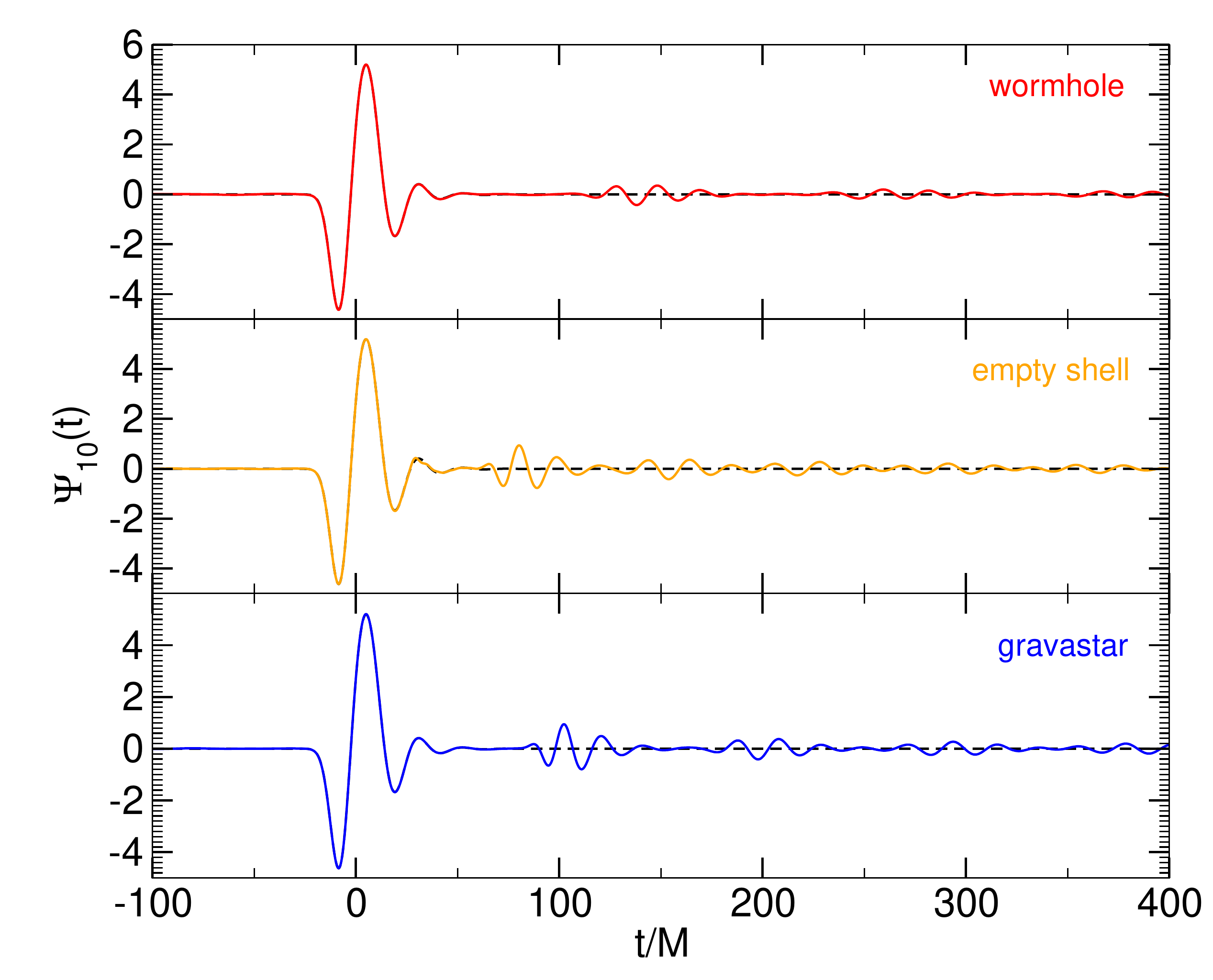}
\includegraphics[width=0.49\textwidth]{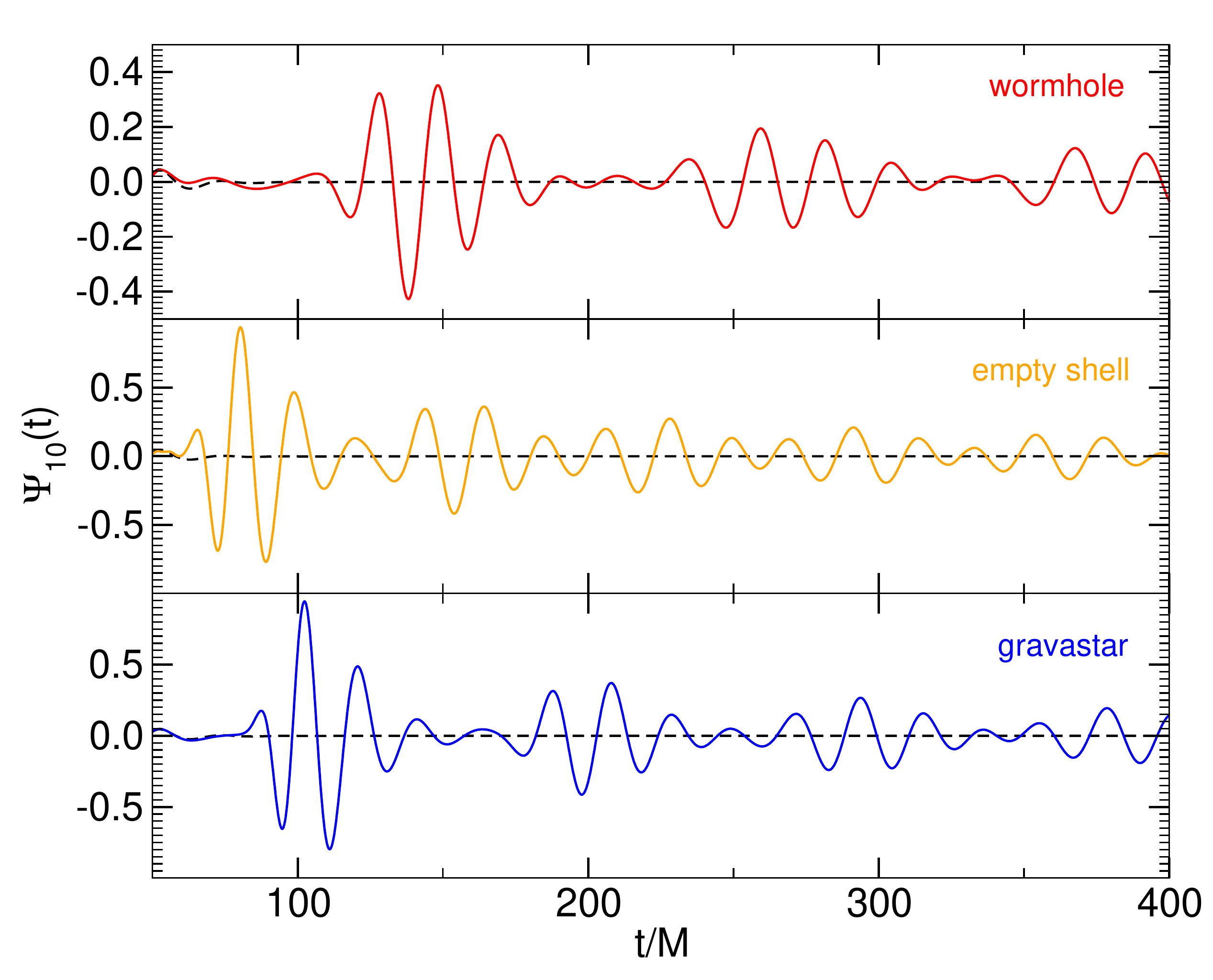}
\caption{Left: A dipolar ($l=1,m=0$) scalar wavepacket scattered off a Schwarzschild BH and off different ECOs with $\ell=10^{-6} M$ ($r_0=2.000001 M$). The right panel shows the late-time behavior of the waveform.
The result for a wormhole, a gravastar, and a simple empty shell of matter are qualitatively similar and display a series of ``echoes'' which are modulated in amplitude and distorted in frequency. For this compactness, the delay time in Eq.~\eqref{Deltat} reads $\Delta t\approx 110 M$ for wormholes,
$\Delta t\approx 82 M$ for gravastars, and $\Delta t\approx 55 M$ for empty shells, respectively.
\label{fig:scattering}}
\end{center}
\end{figure*}

We solve Eq.~\eqref{eqn:TDmastereqn} with initial conditions
\be
\frac{\partial \Psi_{lm}}{\partial t}(0,r)=e^{-(r_*-r_g)^2/\sigma^2}\,,\qquad \Psi_{lm}(0,r)=0\,.
\ee
%
%subjected to initial condition in the frequency domain by performing a Laplace transform and initial conditions. As a source term, we consider a %Gaussian wavepacket
%\begin{equation}
% \tilde S_{lm}= \sqrt{F(r)B(r)}\exp\left[-(r_*-r_g)^2/\sigma^2\right]\,,
%\end{equation}
%%%%
%\pp{pls check} where $S_{lm}:=(2\pi)^{-1}\int_{-\infty}^{\infty}d\omega \tilde S_{lm} e^{-i\omega t}$. 
The waveform obtained by solving Eq.~\eqref{eqn:TDmastereqn} with $r_g=10M,\,\sigma=6M$ is shown in Fig.~\ref{fig:scattering} for various models and compared to the BH case.

As discussed in Ref.~\cite{Cardoso:2016rao}, the initial ringdown signal is basically identical to that of a BH. This part of the signal corresponds to unstable modes at the outermost PS, which is present when $r_0<3M$, and which are associated with the maximum of the potential $V_l$ near $r\sim 3M$. The outermost PS is typically unstable on short timescales, explaining the rapid damping of this {\it ringdown} stage.

On the other hand, when the Schwarzschild horizon is replaced by a surface (as, e.g., in the gravastar case) or by a throat (as in the wormhole case), the potential also develops a minimum 
(i.e, an innermost {\it stable} PS) which can trap low-frequency modes~\cite{1991RSPSA.434..449C,Chandrasekhar:1992ey,Abramowicz:1997qk,Macedo:2013jja,Cardoso:2016rao} (cf. Fig.~\ref{fig:potential}). This inner PS can also be thought of as being caused by the centrifugal barrier, and it may become nonlinearly unstable~\cite{Cardoso:2016rao}. These modes make their way to the waveforms in Fig.~\ref{fig:scattering} in the form of ``echoes'' of the initial PS modes after they leak through the potential barrier: the radiation pulse generated at the potential barrier peak (the PS modes) is then trapped in a semi-permeable cavity bounded between the two PSs.
Indeed, the time delay between two consecutive echoes is roughly the time that light takes for a round trip between the potential barrier. In general, this delay time reads
%%%
\begin{equation}
 \Delta t\sim2\int_{r_{\rm min}}^{3M} \frac{d r}{\sqrt{F B}}\,, \label{Deltat0}
\end{equation}
%%%
where $r_{\rm min}$ is the location of the minimum of the potential shown in Fig.~\ref{fig:potential}. If we consider a microscopic correction at the horizon scale ($\ell\ll M$), then the main contribution to the time delay comes near the radius of the star and therefore,
\begin{equation}
 \Delta t\sim -nM \log\left(\frac{\ell}{M}\right)\,, \quad \ell\ll M\,,  \label{Deltat}
\end{equation}
where $n$ is a factor of order unity that takes into account the structure of the objects.
For wormholes, $n=8$ to account for the fact that the signal is reflected by the two maxima in Fig.~\ref{fig:potential}, whereas for our thin-shell gravastar model and the empty-shell model it is easy to check that $n=6$ and $n=4$, respectively. 
The results shown in Fig.~\ref{fig:scattering} for $\ell=10^{-6} M$ are perfectly consistent with this picture, with the wormhole case displaying longer echo delays than the other cases with the same compactness. Our results show that the dependence on $\ell$ is indeed logarithmically for all the ECOs we studied.

As argued in Ref.~\cite{Cardoso:2016rao}, the logarithmic dependence displayed in Eq.~\eqref{Deltat} implies that even Planckian corrections ($\ell\approx L_P=2\times 10^{-33}\,{\rm cm}$) appear relatively soon after the main burst of radiation, so they might leave an observable imprint in the GW signal at late times. From Eq.~\eqref{Deltat}, a typical time delay reads
%%%
\begin{equation}
 \Delta t\sim 54 (n/4)\,M_{30}\left[1-0.01\log\left(\frac{\ell/L_P}{M_{30}}\right)\right]\, {\rm ms}\,,
\end{equation}
%%%
where $M_{30}:=M/(30 M_\odot)$.

The picture of GW signal scattered off the potential barrier is also supported by two further features shown in Fig.~\ref{fig:scattering}, namely the \emph{modulation} and the \emph{distortion} of the echo signal. In general, modulation is due to the slow leaking of the echo modes, which contain less energy than the initial one. In the wormhole case, this effect is stronger due to the fact that modes can also leak to the ``other universe'' through tunneling at the second peak of the potential. While the amplitude of the echoes is model-dependent, for a given model it depends only mildly on $\ell$.
Distortion is also due to the potential barrier, which acts as a low-pass filter and reflects only the low-frequency, quasibound echo modes. This implies that each echo is a low-frequency filtered version of the previous one and the original shape of the mode gets quickly washed out after a few echoes\footnote{Incidentally, we note that all these features (namely time delay, echoes, modulation, and high-frequency filtering) are precisely what one would expect by the scattering of sound waves in a finite-size cavity.}.
\begin{figure*}[th]
\center
\includegraphics[width=0.49\textwidth]{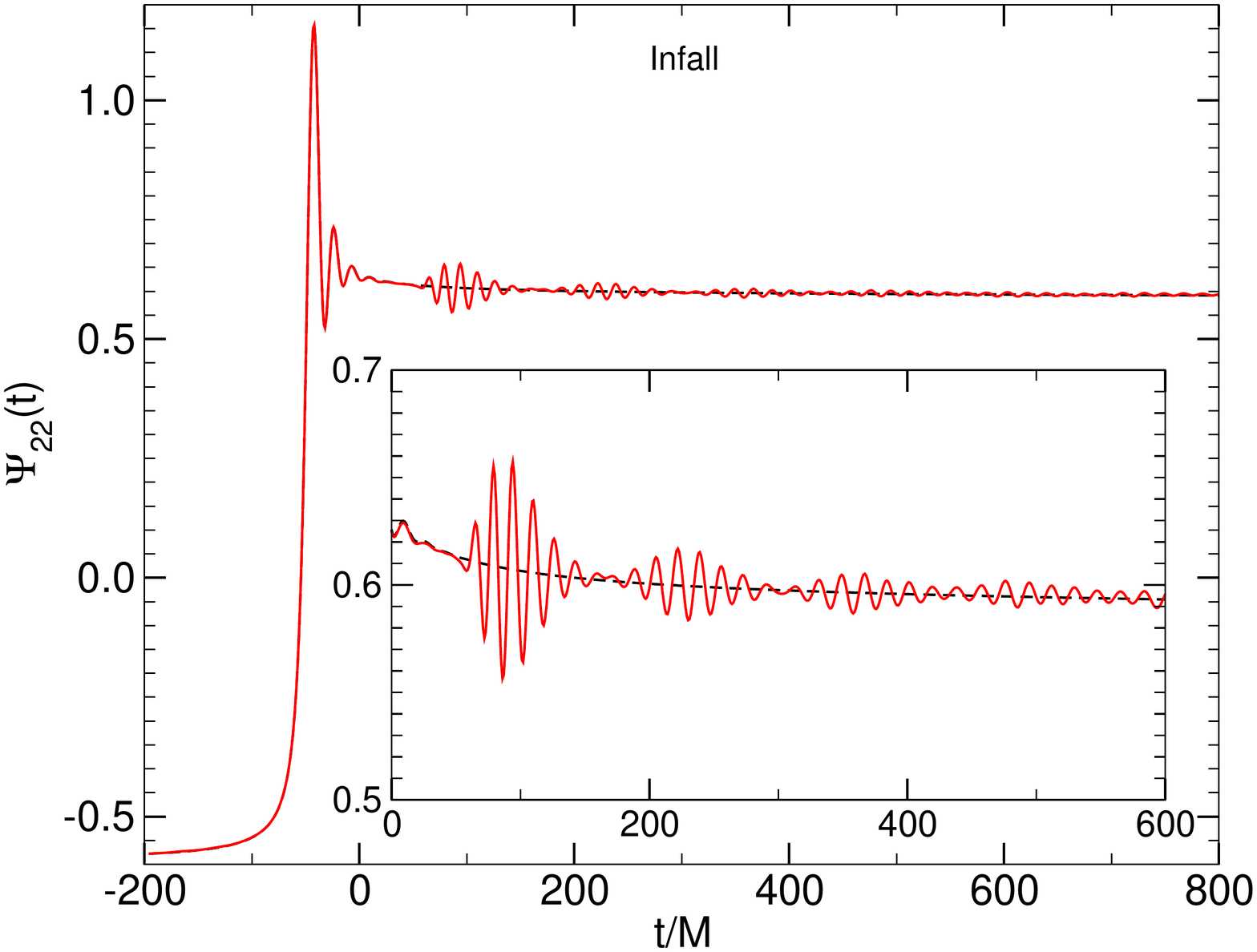}
\includegraphics[width=0.49\textwidth]{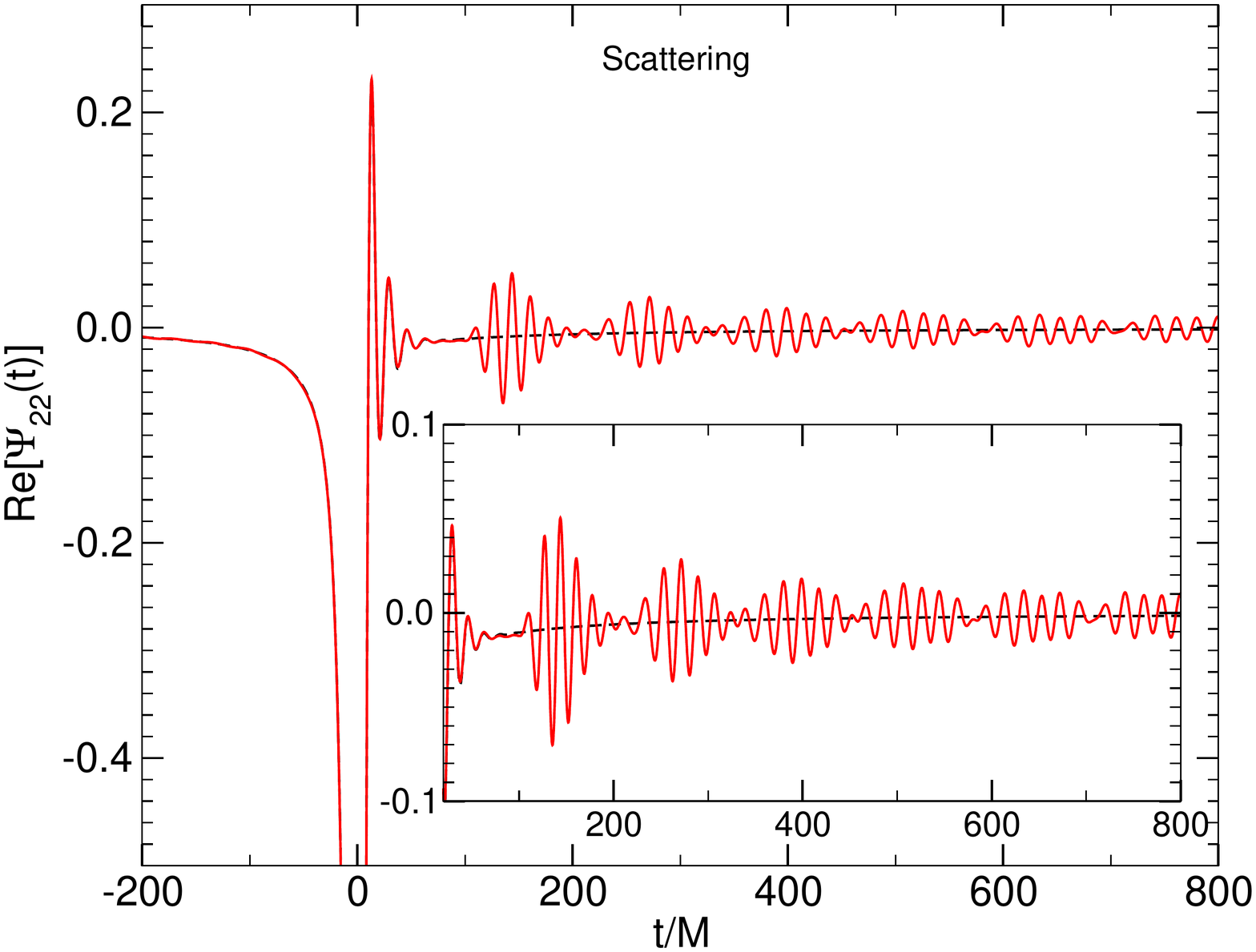}
\caption{Left panel: The waveform for the radial infall of a particle with specific energy $E = 1.5$ into a wormhole with $\ell=10^{-6} M$, compared to the BH case. The BH ringdown, caused by oscillations of the outer PS as the particle crosses through, are also present in the wormhole waveform. A part of this pulse travels inwards and is absorbed by the event horizon (for BHs)
or then bounces off the inner (centrifugal or PS) barrier for ECOs, giving rise to echoes of the initial pulse. This is a low-pass cavity which cleans the pulse of high-frequency components. At late times, only a lower frequency, long-lived signal is present, well described by the QNMs of the ECO.
Right panel: the same for a scattering trajectory, with pericenter $r_{\rm min} = 4.3M$, off a wormhole with 
$\ell= 10^{-6} M$.
The main pulse is generated now through the bremsstrahlung radiation emitted as the particle approaches the pericenter.
The remaining main features are as before.
We show only the real part of the waveform, the imaginary part displays the same qualitative behavior. 
%
%Right panel: same process for different ECOs: wormhole (WH), Florides' fluid star (FS), and gravastar (GS). 
% The particle has specific energy $E=1.5$ and pericenter $r_{\rm min}=4.3M$. 
% The early-time response in \textit{all cases} resembles that of a BH. 
% % The early-time response comes basically from the potential barrier, peaked at $r=3M$.
% The late-time response (shown in the inset) depends on the nature of the object. While in the BH case the wave that passes the barrier is absorbed, in ECOs the wave can be reflected back bringing signatures of the objects' interior. We show only the real part of the waveform, the imaginary part displays the same qualitative behavior. \pp{Discuss how to make plots uniform}
}
\label{fig:funceval}
\end{figure*}
%
%%%%%%%%%%%%%%%%%%%%%%%%%%%%%%%%%%%%%%%%%%%%%%%%%%%%%%%%%%%%%%%%
\subsection{Waves generated by infalling or scattered particles}
%%%%%%%%%%%%%%%%%%%%%%%%%%%%%%%%%%%%%%%%%%%%%%%%%%%%%%%%%%%%%%%%

The features above are observed in a simple scattering process, but are also evident in the GW signal produced by head-on collisions or close encounters, in the test-particle limit. The latter differ from the radial plunge studied in Ref.~\cite{Cardoso:2016rao} in that their pericenter $r_{\rm min}>3M$, i.e. the particle does not cross the radius of the PS (in fact, scattered particles in the Schwarzschild geometry can never get inside the $r=4M$ surface). In order to compute the GW signal, we use the Regge-Wheeler-Zerilli decomposition reviewed in Appendix~\ref{app:RWZ} (cf. Ref.~\cite{Hopper:2010uv} for details). 

We have studied the GW emitted during collisions or scatters between point particles and ECOs; again the general qualitative features are the same as those discussed in Section~\ref{sec:ECO}
and independent of the nature of the ECO. To be specific, we show in Fig.~\ref{fig:funceval} the Zerilli wavefunction for a point particle plunging into (left panel) or scattering off a wormhole
with $\ell=10^{-6}M$, with initial Lorentz boost $E=1.5$. The coordinate system we use is such that the particles are moving along the equator, and it differs - by a $\pi/2$ rotation - from the coordinate axis used in Ref.~\cite{Cardoso:2016rao}. As such, the $l=2$ Zerilli-Moncrief wavefunction, for example, has contributions from azimuthal numbers $m=0,\pm 2$. Note also that it is easy to express these results in a rotated frame~\cite{Palenzuela:2006wp,Gualtieri:2008ux}, and we checked that the waveforms agree up to numerical errors with our previous study~\cite{Cardoso:2016rao}~\footnote{Note however the following typo in the original publication: the bottom right panel of Fig.4 in Ref.~\cite{Cardoso:2016rao} refers to a Lorentz factor $E=1.01$ and not to $E=1.5$ as reported in the paper. This has since been corrected in an Erratum.}.

The left panel of Fig.~\ref{fig:funceval} shows the $l=m=2$ GW waveform generated by point particle plunging radially into a BH and a wormhole with $\ell=10^{-6}M$. The main features were already shown before~\cite{Cardoso:2016rao}, but they are much clearer here: as the particle crosses the outer PS, a pulse of radiation is emitted. This pulse, identical for all ultracompact ECOs, has both an outgoing and ingoing component.
When the central object is a BH, the ingoing pulse disappears towards the event horizon and the outgoing pulse is all that an outside observer receives.
When the central object is an ECO, the ingoing pulse is now trapped between the outer and inner PSs and bounces back and forth,
``echoing'' through the cavity. Because the pulse is of relatively high frequency, there is leakage at each bouncing, and outside observers are able to detect many echoes. At late times the echo acquires a smaller frequency component, identical to the QNMs of the ECO.

This same process is triggered by more generic orbits, even by particles which do {\it not} cross the outer PS, but that approach it sufficiently closely.
This time however, it is the bremsstrahlung radiation that excites the outer PS modes. An example is shown in the right panel of Fig.~\ref{fig:funceval} for a
pericenter at $r_{\rm min} = 4.3M$.
The remaining steps are the same as before: echoes are observed, delayed by an amount that depends exclusively on $\ell$, or in other words, on how big the cavity is.
The {\it amplitude} of the echoes does not seem to be sensitive to $\ell$ but only to the details of the process exciting the main pulse.

In other words, the angular momentum of the infalling particle does not influence significantly the echo structure, and we therefore expect it to be a generic feature of ECOs:
even particles on the last stages of merger will excite echoes as the they plunge through the PS.

% \begin{figure*}[th]
% \begin{center}
% \includegraphics[width=\columnwidth]{Plots/polar_waveform_re}
% \includegraphics[width=\columnwidth]{Plots/polar_waveform_re_late}
% % \includegraphics[width=\columnwidth]{Plots/polar_waveform_im}
% % \includegraphics[width=\columnwidth]{Plots/polar_waveform_im_late}
% \caption{
% Waveforms due to a particle scattered by a Schwarzschild BH and different ECOs: wormhole (WH), fluid star (FS), and gravastar (GS). The particle has specific energy $\mathcal{E}=1.5$ and pericenter $r_{\rm min}=4.3M$.
% Left: The early-time response in \textit{all cases} resembles the one of BH. The early-time response comes basically from the potential barrier, peaked at $r=3M$.
% Right: The late-time response depends on the nature of the object. While in the BH case the wave that passes the barrier is absorbed, in ECOs the wave can be reflected back bringing signatures of the objects' interior. We show only the real part of the waveform, the imaginary part displays the same qualitative behavior. \pp{Maybe we can put the right panel as an inset of the left panel}
% \label{fig:polar_wave}}
% \end{center}
% \end{figure*}

%%%
%
\begin{figure}[th]
\begin{center}
\epsfig{file=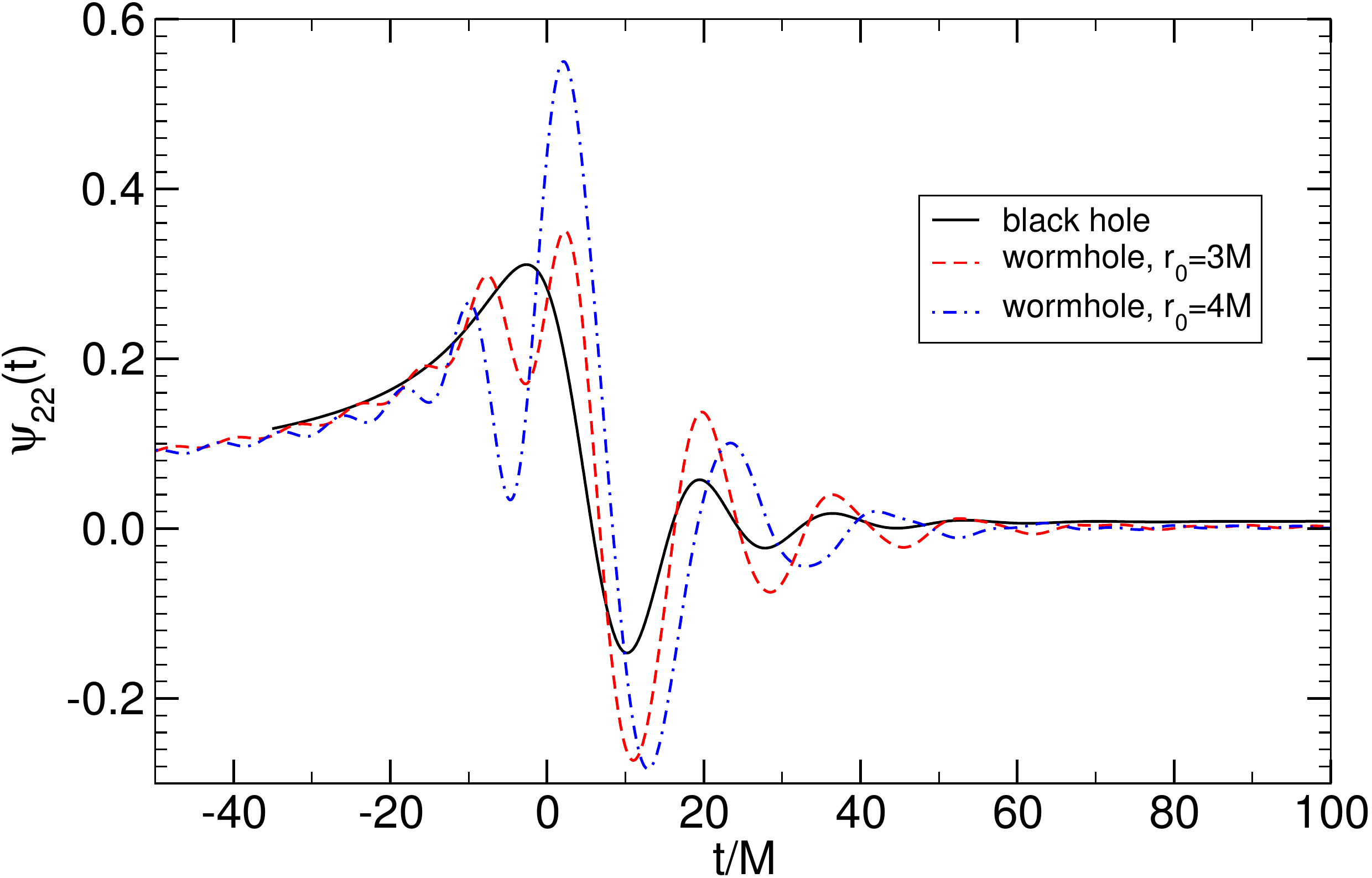,width=0.48\textwidth,angle=0,clip=true}
\caption{GW signal produced by a test particle falling radially into a wormhole with $E=1.01$. We consider the same setup as in Ref.~\cite{Cardoso:2016rao} but for a wormhole without a PS. Without
outer (and inner) PS, the ringdown signal is, clearly, different from that of a BH. Because there is no longer a good resonating cavity, echoes do not appear to be excited. \label{fig:nolightring}}
\end{center}
\end{figure}
Finally, it is clear that a crucial ingredient for the appearance of echoes in the GW signal is the presence of a PS in the spacetime as well as a sufficiently large ``cavity''. As shown in Fig.~\ref{fig:nolightring}, ECOs without a PS display a ringdown different from that of BHs even at early times. Furthermore, because of the absence of trapped states in the spectrum~\cite{Cardoso:2014sna,Cardoso:2016rao} (which, in turn, is due to the absence of a potential well, or an inner PS), the late-time ringdown is simply characterized by a damped sinusoid, without the echo structure. 

%\clearpage
%\newpage
%%%%%%%%%%%%%%%%%%%%%%%%%%%%%%%%%%%%%%%%%%%%%%%%%%%
\section{Head-on collisions of compact boson stars}
%%%%%%%%%%%%%%%%%%%%%%%%%%%%%%%%%%%%%%%%%%%%%%%%%%%
%
\begin{figure}[th]
\begin{center}
\epsfig{file=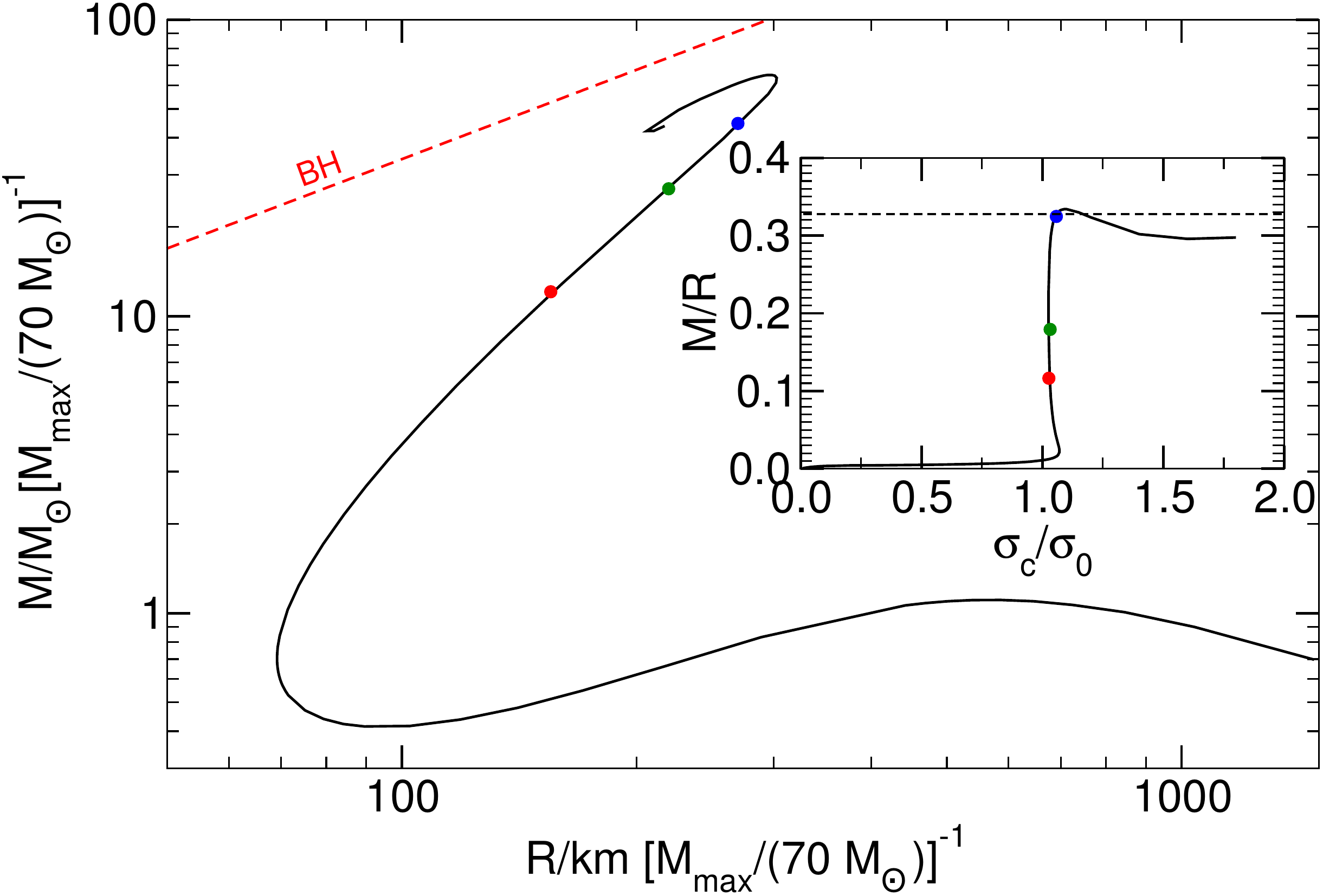,width=0.48\textwidth,angle=0,clip=true}
\caption{Mass-radius relation for a solitonic BS with $\sigma_0=0.05m_P$. In the inset we show the compactness as a function of the central scalar field, $\sigma_c\equiv|\Phi(r=0)|$. The red and the green markers correspond, respectively, to the light BSs with $M/R\approx 0.118$ and to the medium-mass one with $M/R\approx 0.184$ whose numerical evolutions are discussed in the main text. The blue marker indicates a stable BS with nearly maximum mass and  $M/R\approx 1/3$.
The horizontal line in the right panel denotes the compactness of the Schwarzschild PS, $M/R=1/3$. 
\label{fig:BS}}
\end{center}
\end{figure}
In this section we go beyond the point-particle limit previously considered to study the head-collision of two equal-mass ultracompact objects which are initially at rest. First numerical studies in this scenario~\cite{1999PhDT........44B,2005PhDT.........2L} already showed the solitonic behavior of boson stars. Later studies have investigated the outcome of highly relativistic collisions between BSs~\cite{Choptuik:2009ww} in the context of the so-called Hoop Conjecture (crucial for the trans-planckian collision problem). The orbiting case was considered within the conformally flat approximation, which neglects gravitational waves, in~\cite{2010PhDT.......390M}.
Studies concerning gravitational waveforms, directly motivated by GW science, include Refs.~\cite{Palenzuela:2006wp,Palenzuela:2007dm}; nevertheless, studies of collisions of BSs aimed at understanding how well their signal can mimic BHs, are missing.
As discussed in the introduction, solitonic BSs are a natural candidate for this purpose. 

BSs are equilibrium, self-gravitating solutions of the Einstein-Klein-Gordon theory with a minimally-coupled, complex scalar field (cf. Ref.~\cite{Liebling:2012fv} for a review),
\be
S=\int d^4 x \sqrt{-g} \left[\frac{R}{16\pi} -g^{ab}\partial_a\Phi^*\partial_b\Phi-V(|\Phi|^2)\right]\,.\label{action}
\ee
The Einstein equations read $G_{ab}=8\pi T_{ab}^\Phi$, with
%
% \begin{equation}
% R_{ab}-\frac{1}{2}g_{ab}R = \kappa\l( T_{ab}^\Phi + T_{ab}^{\rm matter} \r)\,,\label{eineq}\\
% \end{equation}
% %
% where
%
\be
T^\Phi_{ab}=2\partial_{(a}\Phi^*\partial_{b)}\Phi-g_{ab}\l[\partial^c\Phi^*\partial_c\Phi+V(|\Phi|^2)\r]\,,
\ee
%
% is the energy-momentum of the scalar field. 
whereas the Klein-Gordon equation is $\square\Phi=\frac{d V}{d|\Phi|^2}\Phi$, together with its complex conjugate.
% %
% \begin{equation}
%  \frac{1}{\sqrt{-g}}\pa_a \l(\sqrt{-g}g^{ab}\pa_b\Phi\r)=\frac{d V}{d|\Phi|^2}\Phi\,,\label{eq:phieq}
% \end{equation}
% %
% together with its complex conjugate.
%
We consider solitonic BSs supported by the self-interacting potential~\cite{Friedberg:1986tq,Lee:1991ax}
\begin{equation}
V(|\Phi|^2)=\mu^2|\Phi|^2 (1-2|\Phi|^2/\sigma_0^2)^2\,. \label{potsolitonic BSs}
\end{equation}
The mass $m_{S}$ of the scalar is related to the mass parameter $\mu$ above through $\mu=m_{S}/\hbar$.
Here, $\sigma_0$ is a constant, generically assumed to be of the same order as $\mu$. This is the simplest potential that can support, in the absence of gravity, nontopological solitonic solutions~\cite{Friedberg:1986tq,Lee:1991ax}.
% , i.e., nondispersive scalar field solutions.

%
\begin{figure*}[th]
\begin{center}
\epsfig{file=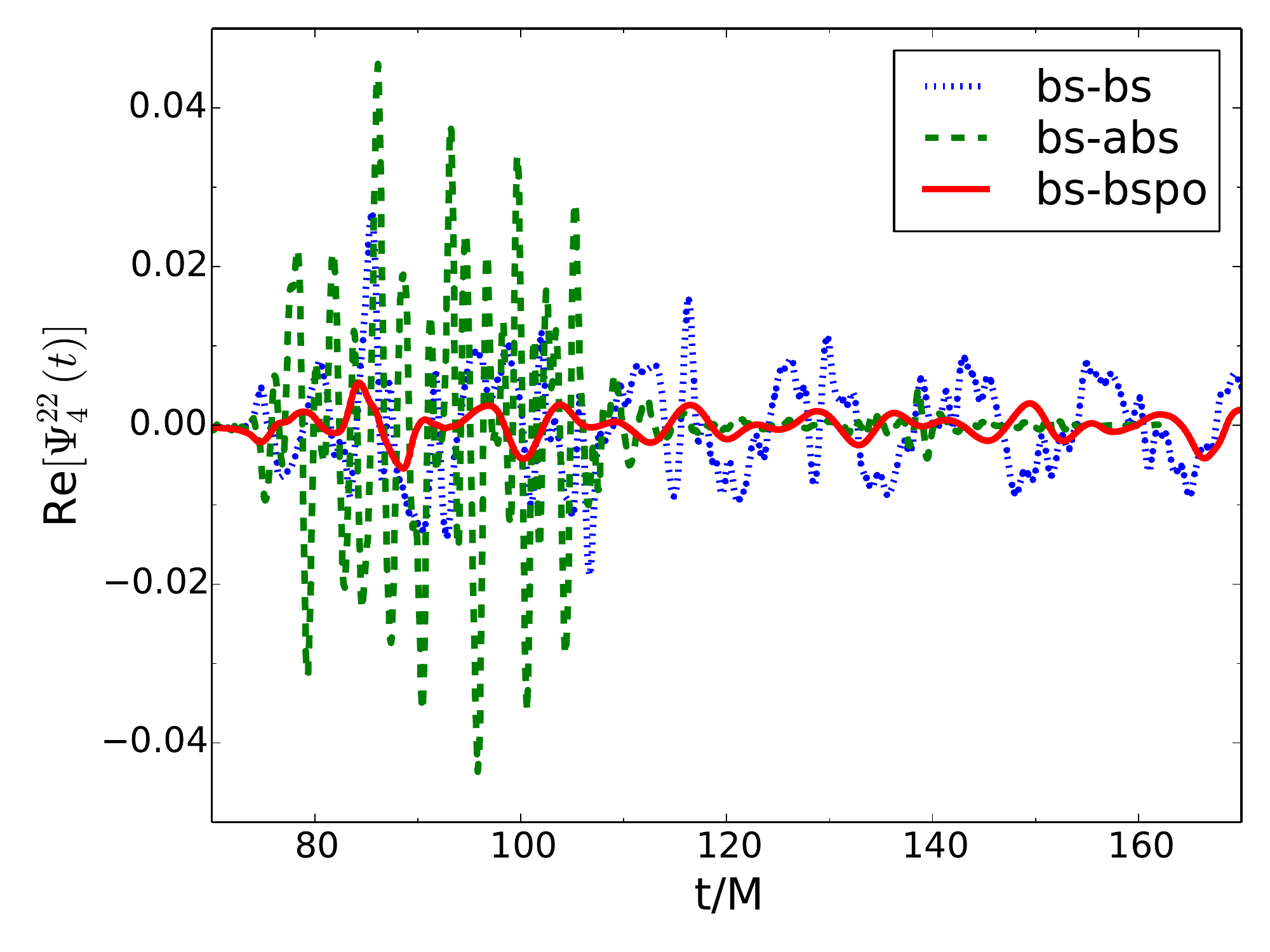,width=0.48\textwidth,angle=0,clip=true}
\epsfig{file=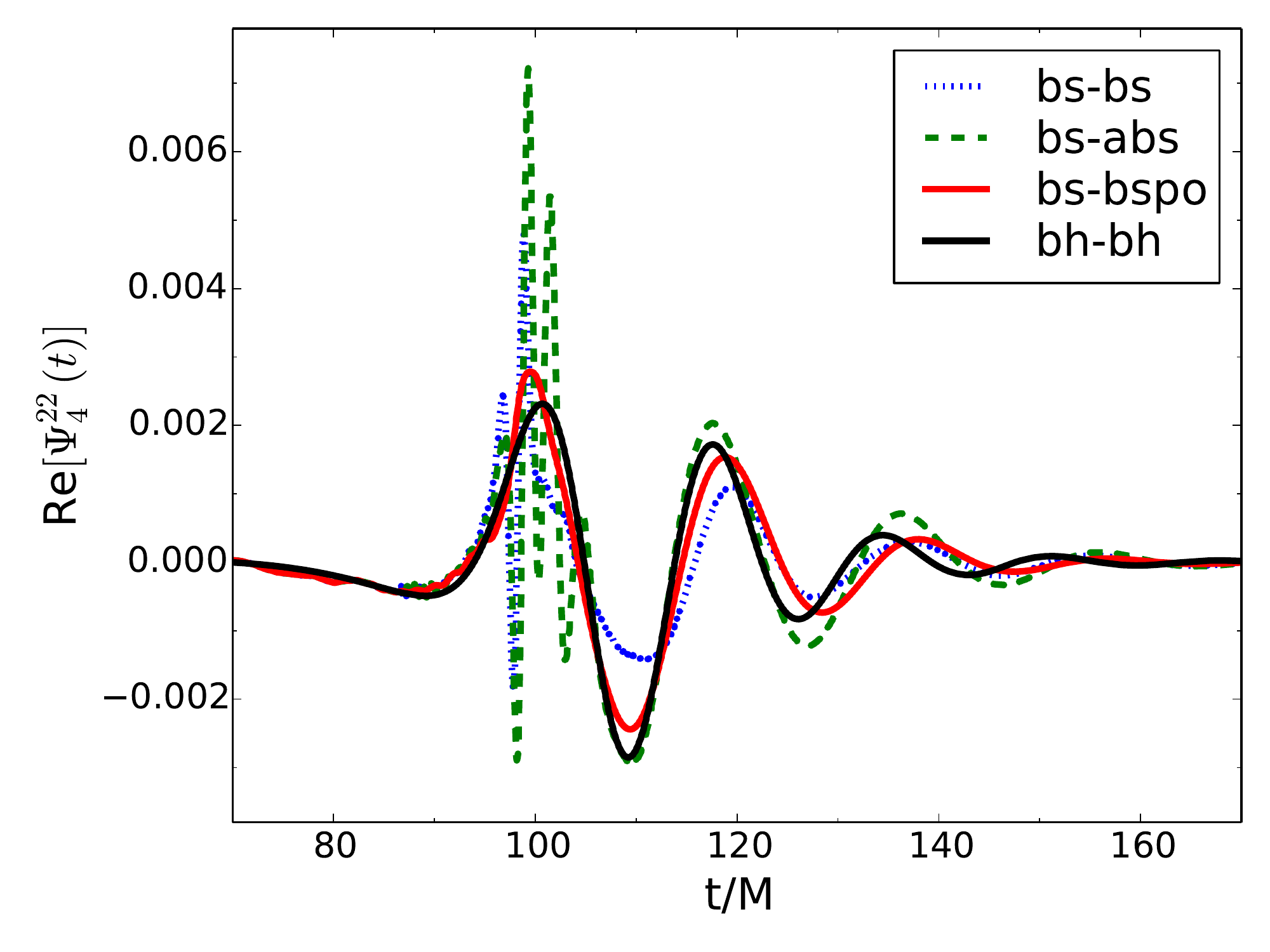,width=0.48\textwidth,angle=0,clip=true}
\caption{ 
GW (i.e., represented by the $l=m=2$ mode of the Newman-Penrose 
$\Psi_4$ scalar), as a function of time, emitted during the head-on collision of two solitonic BSs. Left panel {\em  (low mass)}: The final object is not massive enough to collapse to a BH, so the final fate of the system will depend on the BS configuration; a perturbed BS (bs-bs), annihilation of the stars (bs-abs) or two individual stars after multiple inelastic collisions (bs-bsop).
Right panel {(\em medium mass)}: The final object promptly collapses to a BH, although previously --for some of the configurations, i.e., the bs-bs and the bs-abs-- there is a signature on the GWs produced by the scalar field interaction.}
\label{fig:BSwaveform}
\end{center}
\end{figure*}
Solitonic BSs can be very compact, with the minimum radius of stable, spherically-symmetric configurations comparable to the radius of the PS~\footnote{Strictly speaking, BSs extend all the way to infinity; however, the scalar energy density decreases exponentially at large distances and it is standard practice - which we follow - to define its radius as the point at which $99\%$ of the BS mass is contained.}, i.e.~$R\approx 3M$~\cite{Macedo:2013jja}.
Their maximum mass reads~\cite{Friedberg:1986tq,Kesden:2004qx}
%%%%
\begin{equation}
 M_{\rm max}\sim0.0198\frac{m_P^4}{\mu \sigma_0^2}\,, \label{Mmax}
\end{equation}
%%%%
where the scaling of $M_{\rm max}$ with $\mu$ is exact, while the scaling with $\sigma_0^2$ is only approximate and valid when $\sigma_0\sim\mu\ll m_P$ (with $m_P$ the Planck mass). The field equations for the static, spherically symmetric case are given in Appendix~\ref{app:BSs}, while the numerical setup of the simulations is briefly described in Appendix~\ref{app:numericalcode}. A representative mass-radius relation for solitonic BSs is shown in Fig.~\ref{fig:BS}.

The field equations for the solitonic potential~\eqref{potsolitonic BSs} are stiff, and the scalar field has a very steep profile across a surface layer of thickness $\sim\mu^{-1}$. This stiffness makes the numerical integration particularly challenging. Here, we use the method presented in Ref.~\cite{Macedo:2013jja} to prepare initial data for the spherically symmetric case. The initial state for the binary head-on collision is simply constructed by superposition of two solitonic BSs. The simulations presented below refer to BSs with the same mass and radius, initially at rest and separated by a distance $\approx 2.7 R$;
%\approx (15-22)M$
we have also tried different configurations finding qualitatively similar results.

The theory~\eqref{action} is symmetric under $\Phi\to-\Phi$ and under a $U(1)$ transformation. By using these symmetries, we can straightforwardly change the sign of the Noether charge and/or the phase of the scalar field in either of the two BSs in the initial data. We will present results for three different configurations covering the most extreme interaction dynamics between BSs: (1) a binary of two identical BSs in phase (bs-bs), (2) a binary with a BS and an anti-BS with opposite Noether charge (bs-abs), and (3) a binary with two BSs in phase opposition (bs-bsop) corresponding to a shift of $\pi$ in the phase of one of the stars. Notice that these cases have also been studied in the context of binary mini-BSs~\cite{Palenzuela:2006wp,Palenzuela:2007dm}.

\begin{figure}[th]
\begin{center}
\epsfig{file=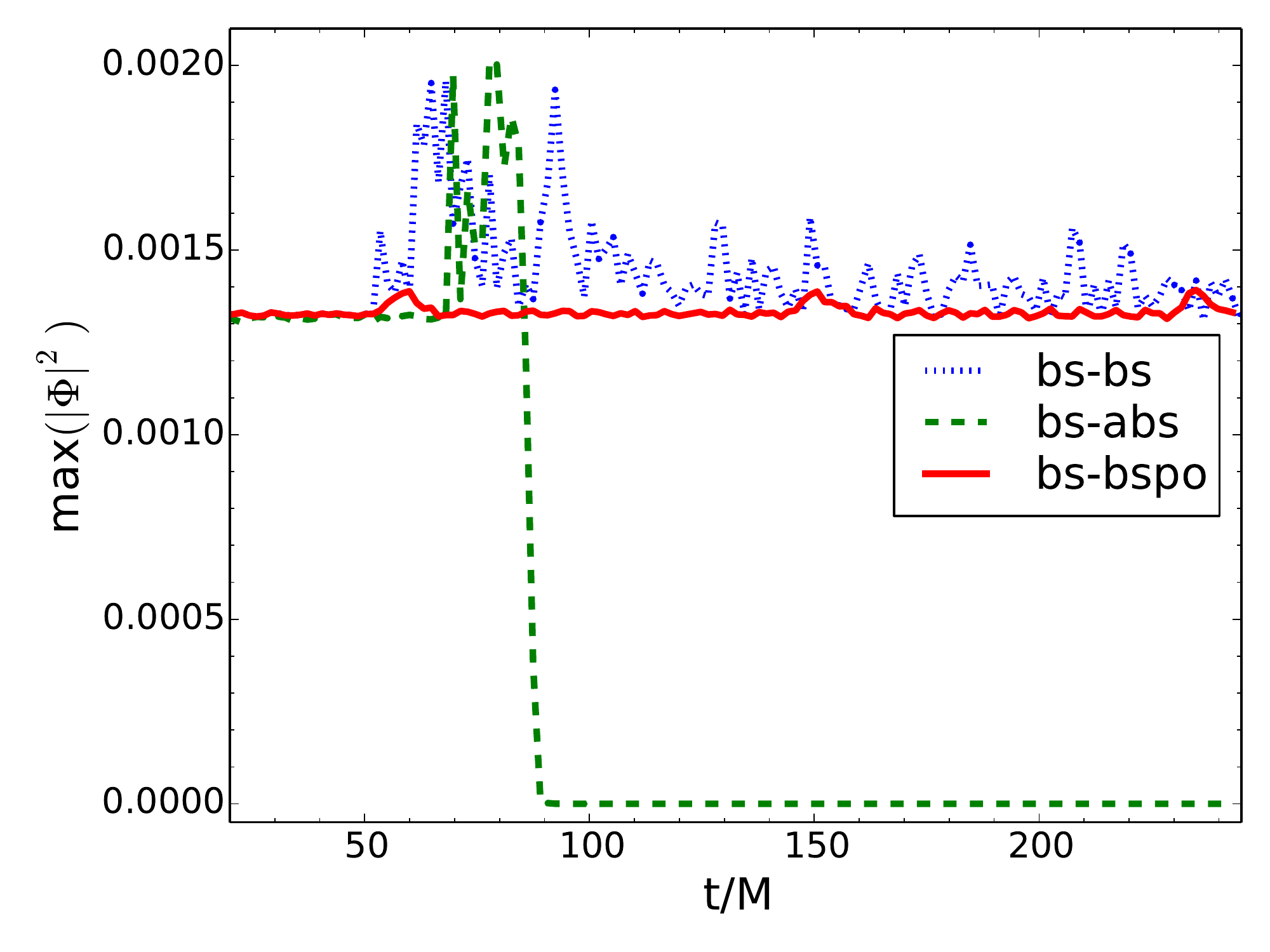,width=0.48\textwidth,angle=0,clip=true}
\caption{Maximum value of the scalar field norm, as a function of time, for the head-on collision of two low-mass solitonic BSs. The bs-bs collision forms, after a relatively long transient, a single perturbed BS. The bs-abs configuration has opposite Noether charges that annihilate soon after they merge, destroying the stars and dispersing/radiating their scalar fields. The
scalar field interaction in the bs-bsop is repulsive and larger than the gravitational attraction, so the system undergoes several inelastic collisions --which compresses the star and leads to small bumps on the scalar field norm that can be observed at $t\approx \{60,160,240\}$-- before
relaxing to a binary at rest with the surfaces barely touching.
\label{fig:BSphi2max}}
\end{center}
\end{figure}

The gravitational waveforms of these three configurations are presented in Fig.~\ref{fig:BSwaveform} for two representative values of the total mass. In the left panel, we consider the evolution of two relatively light BSs with $M/R\approx 0.118$ (red marker in Fig.~\ref{fig:BS}), whose head-on collision could form in principle a BS with nearly maximum mass and with $M/R\approx 1/3$ (blue marker in Fig.~\ref{fig:BS}). In all the cases considered here the waveforms display a qualitative behavior different from that of the head-on collisions of two (non-spinning) BHs with the same masses. 
We argue that this is mainly due to three reasons. First, in order to form a BS which does not exceed the maximum mass in Fig.~\ref{fig:BS}, the compactness of the initial BSs has to be relatively low, so they start merging much before the BH case. 
Secondly, the radius of the final BS (if it eventually forms after a transient stage) roughly coincides with the PS, 
and does not meet the requirements to mimic well a BH ringdown at early times~\cite{Cardoso:2016rao} (as also shown in Fig.~\ref{fig:nolightring}). 
Finally, the scalar fields composing the BSs interact in a non-trivial way during the merger which depends on its specific configuration (i.e., phase shift and Noether charge sign). 
This behavior manifests more clearly in the  maximum of the scalar field norm, which is displayed in Fig.~\ref{fig:BSphi2max}.
The only scenario with a clear merger and relaxation to another BS configuration is the (bs-bs) case. The opposite Noether charges of the (bs-abs) case annihilate during the merger, dispersing and radiating all the scalar field. The (bs-bsop) case is probably the most exotic scenario, since the scalar field interaction induces a repulsive force. Therefore, the stars suffer several inelastic collisions, bouncing back and forth, before losing all their kinetic energy. At late times the two stars are  at rest, next to each other, without merging.

In the right panel of Fig.~\ref{fig:BSwaveform} we show the case of two BSs with moderately high compactness, $M/R\approx 0.184$ (green marker in Fig.~\ref{fig:BS}). For all the configurations considered, the final product of the merger is a Schwarzschild BH. However, the relative phase of the scalar field and the BS charges have again a dramatic impact in the waveform. When the two BSs have initially the same phase, their waveforms are qualitatively different from that of two colliding BHs, regardless the relative sign of the BS charges.
In particular, high frequency oscillations, resulting from the interaction of the BSs scalar fields, appear soon after the merger. The resulting star promptly collapses to a BH, producing the well-known ringdown signal after the merger. However, when the initial BSs have opposite phases the interaction between the scalar fields does not produce any interference pattern and the signal is much more similar to that of a BH. In this latter case, we expect that the small differences arise from the compactness of the initial BSs (which is anyway smaller than the BH case roughly by a factor $2.7$) and by the related fact that the mass of the final BH is not exactly the same in the two cases.

Finally, notice that further increasing the total mass of the system will result in a faster collapse of the final object into a BH, reducing the anomalous signatures of the scalar fields interference on the emitted GWs.

%%%%%%%%%%%%%%%%%%%%%%%%%%%%%%%%%%%
\section{Discussion and Conclusion}
%%%%%%%%%%%%%%%%%%%%%%%%%%%%%%%%%%%

The recent direct detection of GWs~\cite{GW150914,GW151226} has opened two intriguing opportunities related to ECOs, namely constraining these objects as alternatives to BHs and using them as a proxy to probe quantum corrections at the horizon scale. 
In this paper, we have explored some GW signatures that emerge in both scenarios, finding a number of interesting results.
%%%%%%%%%%%%%%%%%%%%%%%%%%%%%%%%%%%
\subsection{Quantum corrections}
%%%%%%%%%%%%%%%%%%%%%%%%%%%%%%%%%%%

First of all, we have extended and clarified the picture proposed in Ref.~\cite{Cardoso:2016rao}. The ringdown signal of an ECO whose compactness is parametrically close to that of a BH displays some universal features. The main burst of radiation during the early-time ringdown phase is only associated with the vibration modes of the PS and does not depend on whether or not the spacetime has a horizon. The pulse of radiation at the PS travels unimpeded towards the event horizon, if the object is a BH. Thus the late stages in the dynamics of BHs are simple. 
By contrast, ECOs have an outer and inner PS: they represent a cavity for perturbations, able to trap them and leak them away on very large timescales.
An outside observer {\it will} see characteristic imprints of the absence of horizons under the form of distorted echoes of the original pulse, which can live
orders of magnitude longer than the timescales usually associated to BHs. 
The scattering of wavepackets or of point particles confirms this picture and shows that the modulation of the ringdown signal is associated with the reflection of the main burst of radiation off the potential barrier at the PS, producing a characteristic train of ``echoes'' at interval $\Delta t$. For a given model, the amplitude of the echoes depends only mildly on the compactness in the $\ell\to 0$ limit. Reflections of the potential barrier give also rise to a distortion of each echo mode, since high-frequency components are filtered out. This is an unfortunate feature for GW searches, because it implies that the signal is quickly washed out after a few echoes. In this context, it will be interesting to find an analytical template for the late-time ringdown waveform, in order to search for these echoes in actual GW data through matched filters. 

It is also interesting to study the echo structure in the presence of rapidly rotating objects. This will presumably give rise to rich structure in the echoes, since the lifetime
of the main burst generated at the PS can be much longer for spinning BHs.

%%%%%%%%%%%%%%%%%%%%%%%%%%%%%%%%%%%
\subsection{Hairy black-holes}
%%%%%%%%%%%%%%%%%%%%%%%%%%%%%%%%%%%
The discussion above also sheds some light on the issue of looking for hairy BHs with GWs~\cite{Cardoso:2016ryw}.
Some - if not all - of these solutions are associated with an extra scale in the problem. For example, a minimally coupled
massive scalar field theory gives rise to non-trivial spinning hairy BHs, which describe a scalar ``cloud'' outside the horizon~\cite{Herdeiro:2014goa}, the spatial extent of which is linked to the mass scale of the field.
What our results teach us is that the GW response of such geometries may be identical - at early times - to those of Kerr,
if their near-horizon geometry is sufficiently close to it: the early time response depends mostly on the properties of the PS. 
Only at late times will the effect of a different geometry or environment become noticeable (see also the overview \cite{Barausse:2014tra}). The lesson is therefore that more sensitive detectors are needed to probe the late-time behavior of the dynamical response of hairy BHs, the bonus being that GR will be also tested during the process.

%%%%%%%%%%%%%%%%%%%%%%%%%%%%%%%%%%%
\subsection{Boson star strawmen}
%%%%%%%%%%%%%%%%%%%%%%%%%%%%%%%%%%%
Having investigated the ringdown signatures of microscopic deviations at the horizon scale, the complementary problem of ECOs as BH mimickers is also interesting. BHs enjoy two remarkable features. Not only does their compactness exceed that of neutron stars, quark stars and BSs, but they also have an arbitrary mass. The former feature implies that two BHs typically merge when they are just a few Schwarzschild radii apart, whereas the latter feature implies that the merger product is a (stable) BH whose mass is roughly the total mass of the binary (modulo GW energy loss).

In light of our results, this simple consideration shows that the limitation of ECOs are more theoretical than phenomenological. Contrived models like wormholes and gravastars can be as massive and compact as BHs and can therefore mimic the inspiral phase up to the merger. However, their dynamics in a comparable-mass, two-body collision and their formation in dynamical processes are difficult to study.
The formation and dynamics of BSs, on the other hand, are relatively simple and well established, but these models are limited by their relatively low maximum compactness and maximum mass. Static BSs seem to be considerably less compact than a Schwarzschild BH, and only very fine-tuned models (marginally) possess a PS~\cite{Macedo:2013jja}. This seems to be associated with the finite compressibility of a scalar field, although it would be interesting to find a general bound ``\`a la Buchdahl''~\cite{PhysRev.116.1027} for generic BSs. Thus, while BSs are viable ECOs, they are of less interest to mimic quantum corrections at the horizon scale and to test the findings of Ref.~\cite{Cardoso:2016rao} in the comparable-mass regime.

Nonetheless, we find that --~in some configurations~-- the collision of solitonic BSs can mimic that of two BHs. In particular, two BSs with opposite phase and relatively large compactness produce an initial waveform which is similar to that of two BHs with the same mass, but they merge to form a BH. On the other hand, if the final object is a BS, the initial compactness needs to be sufficiently small and this produces qualitative differences in the pre-merger waveforms. Furthermore, for generic configurations, the GW signal is markedly different from that of a pair of BHs and even the outcome of the merger does not need to be either a BH or a BS.

A natural extension of our work is to study the quasi-circular coalescence of two solitonic BSs and to check whether the picture that emerges from the head-on collisions remains valid, especially for what concerns the role of the initial phases of the BSs and the existence of configurations that can mimic the entire BH-BH coalescence waveform. In this context, an important discriminator is provided by the tidal deformability that enters the inspiral waveform at fifth post-Newtonian order (cf., e.g, Ref.~\cite{Flanagan:2007ix} and Ref.~\cite{Buonanno:2014aza} for a review). Work on the tidal deformability of BSs is underway (see also the recent Ref.~\cite{Mendes:2016vdr}). For the case of gravastars, the tidal Love numbers vanish in the BH limit~\cite{Pani:2015tga,Uchikata:2016qku} which suggests that some --~but not all~-- models of ECOs can be constrained by a GW measurement of the tidal deformability.

Finally, although the effects discussed here might be rather exotic, we advocate a proactive view: not only has GW astronomy the potential to constrain these models, but there is the exciting prospect for novel, unexpected detections.

%%%%%%%%%%%%%%%%%%%%%%%%%%%%%%%%%%%%%%%%%%%%%%%%%%%%%%%%%%%%%%%%%%%%%%%%%%%%%%
%\noindent{\bf{\em Acknowledgments.}}
%%%%%%%%%%%%%%%%%%%%%%%%%%%%%%%%%%%%%%%%%%%%%%%%%%%%%%%%%%%%%%%%%%%%%%%%%%%%%%
\begin{acknowledgments}
We benefited from discussions with all the participants of the {\it Unifying tests of General Relativity} workshop 
at Caltech. We are grateful to Niayesh Afshordi for useful correspondence and suggestions.
V.C. and S.H. acknowledge financial support provided under the European Union's H2020 ERC Consolidator Grant ``Matter and strong-field gravity: New frontiers in Einstein's theory'' grant agreement no. MaGRaTh--646597. Research at Perimeter Institute is supported by the Government of Canada through Industry Canada and by the Province of Ontario through the Ministry of Economic Development $\&$
Innovation.
C.M. thanks the support from Conselho Nacional de Desenvolvimento Cient\'ifico e Tecnol\'ogico (CNPq).
C.P. acknowledges support from the Spanish Ministry of Education and Science through a Ramon y Cajal grant and from the Spanish Ministry of Economy and Competitiveness grant FPA2013-41042-P. 
This project has received funding from the European Union's Horizon 2020 research and innovation programme under the Marie Sklodowska-Curie grant agreement No 690904 and from FCT-Portugal through the projects IF/00293/2013.
The authors thankfully acknowledge the computer resources, technical expertise and assistance provided by CENTRA/IST. Computations were performed at the clusters
``Baltasar-Sete-S\'ois'' and Marenostrum, and supported by the MaGRaTh--646597 ERC Consolidator Grant.
\end{acknowledgments}
%%%%%%%%%%%%%%%%%%%%%%%%%%%%%%%%%%%%%%%%%%%%%%%%%%%%%%%%%%%%%%%%%%%%%%%%%%%%%%
% \clearpage
% \newpage

\appendix
%%%%%%%%%%%%%%%%%%%%%%%%%%%%%%%%%%%%%%%%%%%%%%%%%%%%%%%%%%%%%
%\section{Oscillating nucleons and DM loss\label{app:thermal}}
%%%%%%%%%%%%%%%%%%%%%%%%%%%%%%%%%%%%%%%%%%%%%%%%%%%%%%%%%%%%%
%

%%%%%%%%%%%%%%%%%%%%%%%%%%%%%%%%%%%%%%%%%%%%%%%
\section{Some models of ECOs} \label{app:models}
%%%%%%%%%%%%%%%%%%%%%%%%%%%%%%%%%%%%%%%%%%%%%%%
In this appendix we provide some details on the models of ECOs used in the main text. All models are spherically symmetric and described by the line element~\eqref{ds2}. Some models are discontinuous across the surface and require a thin shell of matter at $r=r_0$. In this case, Israel's junction conditions~\cite{Israel:1966rt} relate the discontinuities of the extrinsic curvature on the surface with the stress-energy tensor of the thin layer. From these conditions, the surface
energy $\Sigma$ and surface pressure $p$ of the shell read~\cite{Visser:2003ge}
\begin{equation}
\llbracket \sqrt{B}\rrbracket =-4\pi R \Sigma\,,\quad 
\left\llbracket {F'\sqrt{B}}/{F} \right\rrbracket = 8\pi (\Sigma+2p)\,, \label{eq:junction}
\end{equation}
where the symbol
$\llbracket A(R) \rrbracket \equiv \lim_{\epsilon\to0} [A(R+\epsilon)-A(R-\epsilon)]$ denotes the discontinuity of a generic function $A(r)$ across
the shell. 

%%%%%%%%%%%%%%%%%%%%%%%%%%%%%%%%%%%%%%%%%%
\subsection{A toy model: empty thin shell}
%%%%%%%%%%%%%%%%%%%%%%%%%%%%%%%%%%%%%%%%%%
The simplest model that displays ringdown echoes is an empty thin shell of matter located at $r=r_0$. The line element is Eq.~\eqref{ds2} with
%%%
\begin{equation}
 F=B=\left\{\begin{array}{l}
             1-2M/r\qquad r>r_0 \\
             1 \qquad \hspace{1.3cm} r<r_0
            \end{array}\right.\,,
\end{equation}
%%%
Equations~\eqref{eq:junction} imply $\Sigma>0$ for any $r_0>2M$, whereas the dominant energy condition on the shell, $|p|\leq \sigma$, implies $r_0\geq 25 M/12\approx 2.08 M$. Note that this solution is unstable against radial perturbations when $r_0\lesssim 2.37 M$~\cite{PhysRevD.44.1891}.

In order to avoid a discontinuity of the metric at the shell, here we consider a slightly different model in which the metric is smooth everywhere~\cite{Barausse:2014tra}. We take the ansatz~\eqref{ds2} with $F(r)=B(r)=1-2m(r)/r$ and 
\beq
F(r)=B(r)=1-\frac{M}{r}\left(1 + {\rm erf}\left[(r-r_0)/L\right]\right)\,,
%
%m(r)&=&\frac{M}{2}\left(1 + {\rm erf}\left[\frac{r-2M}{\ell}\right]\right)\,,\label{bumpyII}
\eeq
where erf is the error function. This metric describes matter fields localized at $r_0$, which we take to be $r_0=2M$ with a spatial extent $L$. The metric above corresponds to a particular case of model~II of matter-bumby BHs studied in Ref.~\cite{Barausse:2014tra}.

%%%%%%%%%%%%%%%%%%%%%%%%%%%%%%%%%%%%%%%%%%%%%%
\subsection{Thin-shell, traversable wormholes}
%%%%%%%%%%%%%%%%%%%%%%%%%%%%%%%%%%%%%%%%%%%%%%
\label{sec:wormhole}
We consider the same model of a traversable wormhole~\cite{Morris:1988tu,VisserBook} used in Ref.~\cite{Cardoso:2016rao}, which is obtained by identifying two Schwarzschild metrics with the same mass $M$ at the throat $r=r_0>2M$. In Schwarzschild coordinates, the two metrics are identical and described by Eq.~\eqref{ds2} with $F=B=1-2M/r$. Because Schwarzschild's coordinates do not extend to $r<2M$, we use the tortoise coordinate $dr/dr_*=\pm F$, where the upper and lower signs refer to the two different universes connected at the throat at $r_0=0$. The surgery at the throat requires a thin shell of matter with surface density and surface pressure~\cite{VisserBook}
\begin{equation}
\Sigma =-\frac{\sqrt{1-2M/r_0}}{2\pi r_0}\,, \quad p =\frac{1}{4\pi r_0} \frac{(1-M/r_0)}{\sqrt{1-2M/r_0}}\,,
\end{equation}
respectively. The weak energy condition is violated ($\Sigma<0$), whereas the strong and null energy conditions are satisfied when the throat is within the PS, $r_0<3M$.

%%%%%%%%%%%%%%%%%%%%%%%%%%%%%%%%%%%
\subsection{Thin-shell gravastars}
%%%%%%%%%%%%%%%%%%%%%%%%%%%%%%%%%%
We consider the thin-shell gravastar model~\cite{Visser:2003ge} studied, e.g., in Ref.~\cite{Pani:2009ss}, which is described by the line element~\eqref{ds2} with
%%%

%%%
\begin{equation}
 F=B=\left\{\begin{array}{l}
             1-2M/r\qquad \hspace{0cm}r>r_0 \\
             1 -\Lambda r^2/3\qquad \hspace{0cm} r<r_0
            \end{array}\right.\,,
\end{equation}
%%%
when $\Lambda=6 M/r_0^3$, both $F$ and $B$ are continuous across the shell (more generic, thin-shell gravastar models have been recently studied in Ref.~\cite{Uchikata:2016qku}). Note that the empty shell model discussed above is a particular case of this gravastar model when $\Lambda=0$.
Although $F$ and $B$ are continuous, their derivatives are not and this requires a thin shell with vanishing energy density and negative surface pressure~\cite{Pani:2015tga}.

% %%%%%%%%%%%%%%%%%%%%
% \subsection{Florides' fluid stars}
% %%%%%%%%%%%%%%%%%%%%
% 
% \pp{Are we going to show these?}

%%%%%%%%%%%%%%%%%%%%%%%%%%%%%%%%%%%%%%%%%%%%%%
\subsection{Boundary conditions at the shell}
%%%%%%%%%%%%%%%%%%%%%%%%%%%%%%%%%%%%%%%%%%%%%%
Some of the models presented above require thin shells of matter across which the metric functions are discontinuous. Gravitational perturbations of discontinuous geometries can be studied by using the thin-shell formalism developed in Ref.~\cite{Pani:2009ss}. In the main text we considered the scattering of scalar wavepackets, for which the junction conditions are easier to obtain.

In the frequency domain (FD), the Klein-Gordon equation on the background~\eqref{ds2} can be written as
%%%
\begin{align}
\left[\frac{d^2}{d r_*^2} +\omega^2- V_l(r)\right]X_{l m\o}(r) = 0\,, \label{waveeq}
\end{align}
%%%%
with $dr/dr_*=\sqrt{FB}$ and
%%%%
\begin{equation}
V_l(r)=F\left(\frac{l(l+1)}{r^2}+\frac{B'}{2r}+\frac{B F'}{2r F}\right)\,.
\end{equation}
%%%%
Here, primes stand for derivative with respect to $r$.
Therefore, the potential is generically discontinuous if $F$, $B$, or their derivatives have a jump across the shell.
Given that Eq.~\eqref{waveeq} is homogeneous, we can assume that $X_{lm\omega}$ is continuous without loss of generality. In this case, even if $V_l$ has some finite jump, the integral of $V_l X_{lm\omega}$ across the shell is continuous. Therefore, by integrating all terms in Eq.~\eqref{waveeq} across the shell we obtain the junction condition
%%%
\begin{equation}
 \left\llbracket {d X_{l m\o}(r)}/{dr_*}\right\rrbracket =0\,, \label{BCscalar}
\end{equation}
%%%
i.e., the first derivative (with respect to the tortoise coordinate) of the scalar wavefunction is continuous. However, note that in Schwarzschild coordinates the first derivative might be discontinuous due to the definition of $r(r_*)$. For example, for the wormhole considered above $dr/dr_*=\pm F$ and therefore the first derivative (with respect to the Schwarzschild coordinate) changes sign across the shell. In the main text, we imposed the junction condition~\eqref{BCscalar} when solving Eq.~\eqref{eqn:TDmastereqn}.

A similar procedure was done for gravitational perturbations, where now assumptions on the stiffness of the matter at the shell have to be dealt with. In practice, we implemented a condition akin to~\eqref{BCscalar} for the Zerilli-Moncrief wavefunction~\cite{Cardoso:2016rao}.
%%%%%%%%%%%%%%%%%%%%%%%%%%%%%%%%%%%%%%%%%%%%%%%%%%%%%%%%%%%
\section{Regge-Wheeler-Zerilli formalism} \label{app:RWZ}
%%%%%%%%%%%%%%%%%%%%%%%%%%%%%%%%%%%%%%%%%%%%%%%%%%%%%%%%%%%

Our numerical perturbation theory results shown in Fig.~\ref{fig:funceval}
are found by solving the first-order field equations in Regge-Wheeler gauge
using a FD code. In what follows
one would typically use Schwarzschild $r$ to express spatial dependence.
However, given that in this work we consider the wormhole model
in addition to a Schwarzschild background, 
we make use the tortoise coordinate, as explained in Sec.~\ref{sec:wormhole}.
In the both the wormhole and BH cases $r_* \to \infty$ as $r \to \infty$
in the primary universe. On the other hand, 
in the BH case $r_* \to -\infty$ as $r \to 2M$, while
for the wormhole model $r_* \to -\infty$ corresponds $r \to \infty$ in
the other universe.
%Our infall results follow from an equivalent
%procedure to that described in Ref.~\cite{Cardoso:2016rao} (although
%there the particle was assumed to fall on the pole,
%whereas here it is confined to the equator). For scattering,
%the particle never leaves the first universe, and so the only
%difference between the BH and wormhole cases is the boundary
%conditions set at the horizon (BH) or throat (wormhole).
In both cases, for a given radiative $lm$ mode the field equations reduce to a 
single 1+1 wave equation,
\begin{align}
\left[-\frac{\pa^2}{\pa t^2} + \frac{\pa^2}{\pa r_*^2} - V_l(r_*)\right]
\Psi_{l m}(t,r_*) = S_{l m}(t,r_*) .
\label{eqn:TDmastereqnSrc}
\end{align}
Here $V_l (r_*)$ is either the Zerilli potential ($l+m$ even) or the Regge-Wheeler
potential ($l+m$ odd).  
The particle is assumed to be confined to the 
equator $\theta = \pi /2$ and then the 
source contains terms proportional to the Dirac 
delta function and its first derivative
\begin{align}
\label{eqn:sourceTD}
S_{l m}(t,r_*) = 
G_{l m} (t) \, \delta \left( r_* - r_{*p} \right) 
 + F_{l m} (t) \, \delta' \left( r_* - r_{*p} \right) ,
\end{align}
where $r_{*p} = r_{*p} (t)$ is the particle's radial location measured in $r_*$.
The time dependent functions $G_{l m}(t)$ and $F_{l m}(t)$ result from the tensor 
spherical harmonic decomposition of the stress-energy 
tensor of the point mass and subsequently evaluating
$r \rightarrow r_p(t)$ and $\varphi \rightarrow \varphi_p(t)$.  Their specifics 
will be discussed in further detail below.
%(Note that in this Appendix alone, to avoid confusion with the source
%term, $F_{lm}$ we use 
%$f(r) \equiv 1-2M/r$ as the gravitational redshift instead of $F$.)

We move to the FD with a Fourier transform
\begin{align}
\begin{split}
\label{eqn:FourierTransf}
X_{l m\o}(r_*) &= \int_{-\infty}^\infty 
\Psi_{l m}(t,r_*) \, e^{i \o t}  dt , 
\\
Z_{l m\o}(r_*) &= \int_{-\infty}^\infty 
S_{l m}(t,r_*) \, e^{i \o t}  dt ,
\end{split}
\end{align}
and the inverse relations
\begin{align}
\begin{split}
\label{eqn:invFourierTransf}
\Psi_{l m}(t,r_*) &= \frac{1}{2\pi} \int_{-\infty}^\infty 
X_{l m\o}(r_*) \, e^{-i \o t}  d\o , 
\\
S_{l m}(t,r_*) &= \frac{1}{2\pi} \int_{-\infty}^\infty 
Z_{l m\o}(r_*) \, e^{-i \o t}  d\o .
\end{split}
\end{align}
Given these, the master equation \eqref{eqn:TDmastereqnSrc}
 takes on the following FD form
\begin{align}
\label{eqn:FDmastereqn}
\left[\frac{d^2}{dr_*^2} +\o^2 -V_l(r_*)\right]
X_{l m\o}(r_*) = Z_{l m\o}(r_*).
\end{align}
We assume retarded boundary conditions and
asymptotically unit-amplitude homogeneous solutions,
\begin{align}
\hat{X}^\pm_{l m\o}(r_* \to \pm \infty )
= e^{\pm i \omega r_*} .
\end{align}
%In the BH case $+$ corresponds to an outgoing wave at infinity, 
%and $-$ a downgoing wave at the horizon,
%but in the wormhole case $-$ is an outgoing infinity solution in the other universe.
%When integrating through the throat of the wormhole, we apply the junction
%condition described in Sec.~\ref{sec:wormhole}.
The solution to Eq.~\eqref{eqn:FDmastereqn} 
follows from the method of variation of parameters,
\begin{align}
\label{eq:FDInhomog}
X_{l m\o} (r_*) = c^+_{l m\o}(r_*) \, \hat{X}^+_{l m\o}(r_*)
+ c^-_{l m\o}(r_*) \, \hat{X}^-_{l m\o}(r_*) ,
\end{align}
where
\begin{align}
\label{eq:cPM}
\begin{split}
c^+_{lm\o}(r_*) &= \frac{1}{W_{lm\o}} \, \int_{-\infty}^{r_*} 
dr_*' \ \hat{X}^-_{lm\o}(r_*') \, Z_{lm\o}(r_*') , \\
c^-_{lm\o}(r_*) &= \frac{1}{W_{lm\o}} \, \int_{r_*}^{\infty} 
dr_*' \ \hat{X}^+_{lm\o}(r_*') \, Z_{lm\o}(r_*') ,
\end{split}
\end{align}
and $W_{lm\o}$ is the (constant-in-$r_*$) Wronskian.
Extending the integrals in Eq.~\eqref{eq:cPM} over all space provides the 
normalization coefficients
\begin{align}
\label{eq:normC}
C_{l m\o}^{\pm} 
= \frac{1}{W_{l m\o}} \int_{-\infty}^{\infty} dr_*
\ \hat X^{\mp}_{l m\o} (r_*) Z_{l m\o} (r_*).
\end{align}
Finally, inserting the specific form of the source from 
Eqs.~\eqref{eqn:FourierTransf}
and \eqref{eqn:sourceTD} this becomes an integral over time
\cite{Hopper:2010uv}.
%\begin{align}
%\label{eqn:normC}
%\begin{split}
%&C_{lm\omega}^\pm  =  \frac{1}{W_{lm\omega}} \int_{-\infty}^{\infty}
%\Bigg[ 
% \frac{1}{f_{p}} \hat X^\mp_{lm\omega}
% G_{lm}  \\ 
%& \hspace{5ex} + \l( \frac{2M}{r_{p}^2 f_{p}^{2}} \hat X^\mp_{lm\omega}
% - \frac{1}{f_{p}} 
% \frac{d \hat X^\mp_{lm\omega}}{dr} \r) F_{lm}
% \Bigg]  e^{i \omega t}  \, dt .
% \end{split}
%\end{align}
%All $r$-dependent quantities under the integral are evaluated at $r=r_p(t)$.
%This includes the homogeneous solutions $\hat X^\mp_{lm\omega} (r_p)$ and 
%$f_p \equiv f(r_p)$.
While a full solution to Eq.~\eqref{eqn:FDmastereqn} requires the functions
$c^{\pm}_{lm\omega}(r_*)$, the constants $C^\pm_{lm\omega}$ are all that 
are required to
compute the total radiated energy and angular momentum, as well 
as the waveform at infinity.

From a practical standpoint, repeated evaluations of the integral 
\eqref{eq:normC} for a range of $l,m$, and $\omega$ 
make up the brunt of our calculation.
The integral converges if the source coefficients
$G_{lm} (t)$ and $F_{lm} (t)$ die off as $t \to \pm \infty$,
or equivalently, as $r_* \to \pm \infty$. 
These source coefficients are unique to the specific master function used,
and while all master function sources decay
rapidly at the horizon (for the BH case), at infinity they have differing behaviors.
For bound motion it is best to use the
Zerilli-Moncrief \cite{Moncrief:1974am} (ZM) function
(for $l+m$ even) and the Cunningham-Price-Moncrief \cite{Cunningham:1978zfa} 
(CPM) function (for $l+m$ odd)
because they allow for simple time domain reconstruction of the metric 
perturbation amplitudes.
However, for unbound motion these variables are less than ideal.
The ZM source tends to a constant at infinity while
the CPM source falls off slowly as $r_p^{-1}$.
Therefore, in the even-parity sector it is better to use 
Zerilli's original variable \cite{Zerilli:1971wd}, which has a source that decays as 
$r_p^{-1}$. Meanwhile, in the odd-parity sector Regge and Wheeler's \cite{Regge:1957td} 
original variable is preferable,
since its source decays as $r_p^{-3}$. These original variables are essentially
the time derivatives of the ZM and CPM variables, which accounts for the more-rapid
source fall-off. It is possible to take further time derivatives and
define new master functions with source terms that decay even faster.
In particular, the time derivative of the Zerilli function has a source
which falls off as $r_p^{-3}$ at large distance, making it far more efficient
than the Zerilli variable.
In an upcoming work \cite{Hopper:2016} the relation between all these master functions will
be explored and the specific details of their sources will be given.

When using the ZM and CPM variables, 
the total energy radiated for a given $lm$ mode to infinity is
\begin{align}
\label{eqn:enRad}
 E^+_{l m}
= \frac{1}{128 \pi^2} 
\frac{(l+2)!}{(l-2)!}
\int \o^{2}  \left|C^\pm_{l m\o}\right|^2 d\o .
\end{align}
The gauge invariant waveform can also be computed from our code
As $r_* \to \infty$ the FD particular solutions go to
\begin{align}
X^+_{lm\omega} (r_* \to \infty) = C^+_{lm\omega} e^{i \o r_*}
\end{align}
Thus, in order to evaluate these at retarded time $u = t - r_*$ and $r_* \to \infty$
we compute
\begin{align}
\label{eqn:waveform}
\Psi_{l m}(u, r_*\to \infty) 
&= 	
\frac{1}{2\pi} \int_{-\infty}^\infty 
C^+_{l m\o} \, e^{-i \o u}  d\o .
\end{align}
As shown in Ref.~\cite{Martel:2005ir}, these can be summed over $l$ and $m$ 
to form the transverse-traceless metric perturbation.
In practice, the integrals \eqref{eqn:enRad} and \eqref{eqn:waveform}
must be discretized. In order to obtain our results we sampled $\Delta \omega$
as small as $2.5 \times 10^{-5} / M$ in a frequency range
as wide as $-1/M \le \omega \le 1/M$, skipping only
the zero frequency mode which contributes no radiation and provides only a 
constant offset to the waveform.

%%%%%%%%%%%%%%%%%%%%%%%%%%%%%%%%%%%%%%%
\section{Spherically-symmetric BSs} 
\label{app:BSs}
%%%%%%%%%%%%%%%%%%%%%%%%%%%%%%%%%%%%%%%

To describe spherically symmetric BSs we consider the 
line element
\begin{equation}
ds_0^2=-e^{v(r)}dt^2+e^{u(r)}dr^2+r^2 d\Omega^2\,. \label{metric0}
\end{equation}
%%%
and the harmonic ansatz for 
scalar field reads
\begin{equation}
\Phi_0(t,r)\equiv \phi_0(r)e^{-i\omega t}\,,
\end{equation}
where $\phi_0(r)$ is a real function. Although the scalar field is time dependent, the Einstein-Klein-Gordon system admits static and spherically symmetric metrics~\cite{Kaup:1968zz,Ruffini:1969qy,Colpi:1986ye,Gleiser:1988rq,Gleiser:1988ih,Schunck:2003kk,Liebling:2012fv}. With the ansatz above, the 
field equations
% , obtained from \eqref{eineq}--\eqref{eq:phieq}, 
read
\begin{eqnarray}
\frac{1}{r^2}\l(r\,e^{-u}\r)' -\frac{1}{r^2}&=&-8\pi\rho\,,\label{bg11}\\
e^{-u}\l(\frac{v'}{r}+\frac{1}{r^2}\r)-\frac{1}{r^2}&=&8\pi p_{\text{rad}}\,,\label{bg22}\\
\phi_0''+\l(\frac{2}{r}+\frac{v'-u'}{2}\r)\phi_0'&=&e^u\l({U_0}-\omega^2e^{-v}\r)\phi_0\,,\label{bgscalar}
\end{eqnarray}
where a prime denotes the derivative with respect to $r$ and $U_0 \equiv \left.{d V(x)}/{dx}\right|_{x=\phi_0}$. The stress-energy tensor of the scalar field corresponds to that of an anisotropic fluid with  density $\rho$, radial pressure $p_{\text{rad}}$, and tangential pressure $p_{\text{tan}}$ given by 
\begin{eqnarray}
\rho&\equiv&-{{T^\Phi}_t}^t=\omega^2e^{-v}\phi_0^2+e^{-u}(\phi_0')^2+V_0\,,\\
p_{\text{rad}}&\equiv&{{T^\Phi}_r}^r=\omega^2e^{-v}\phi_0^2+e^{-u}(\phi_0')^2-V_0\,,\\
p_{\text{tan}}&\equiv&{{T^\Phi}_\theta}^\theta=\omega^2e^{-v}\phi_0^2-e^{-u}(\phi_0')^2- V_0\,.
\end{eqnarray}
where $V_0=V(\phi_0)$. Unlike the case of perfect fluid stars, the complex scalar field behaves like an anisotropic fluid, $p_{\text{rad}}\neq p_{\text{tan}}$. 
% Equations~\eqref{bg11}--\eqref{bgscalar} can be solved numerically with suitable boundary conditions (see Sec.~\ref{sec:background}) to obtain the background metric and scalar field configuration.

It is convenient to rescale the equations in units of $\Lambda \mu$, with $\Lambda = (8 \pi)^{1/2}\sigma_0$. We use~\cite{Friedberg:1986tq,Kesden:2004qx,Macedo:2013jja}
\begin{eqnarray}
&&r\to \frac{\tilde{r}}{\Lambda \mu},\quad
m(r)\to \frac{\tilde{m}(\tilde{r})}{\Lambda \mu},\nn\\
&&\omega \to \tilde{\omega} \Lambda \mu,\quad
\phi_0 (r) \to \frac{\sigma_0 \tilde{\phi}_0(\tilde{r})}{\sqrt{2}}\,.\nn
\end{eqnarray}
where $m$ is the mass function, defined through $e^{-u}=1-2m/r$. The BS configuration is found by integrating the above equations from the origin, with the boundary conditions $v(0)=v_0$, $\tilde{m}(0)=0$, and $\tilde{\phi}_0(0)=\sigma_c$.  We can perform a time-rescaling, setting $v_0=0$. By imposing that the scalar field goes to zero at infinity and the spacetime tends to the Schwarzschild one, the problem becomes a one-parameter boundary value problem for the frequency $\tilde\omega$. We follow Ref.~\cite{Macedo:2013jja}, using a shooting method to solve the differential equations. Note that, differently from other BS potentials, for solitonic BSs there may be two solutions with the same central field $\sigma_c$, as can be evident from the inset of Fig.~\ref{fig:BS}.

%%%%%%%%%%%%%%%%%%%%%%%%%%%%%%%%%%%%%%%
\section{BSs numerical setup}
\label{app:numericalcode}
%%%%%%%%%%%%%%%%%%%%%%%%%%%%%%%%%%%%%%%

We solve the Einstein-Klein-Gordon equations to describe self-gravitating scalar fields modeling binary BS systems. We use the BSSN formulation \cite{1995PhRvD..52.5428S, 1999PhRvD..59b4007B} to implement the Einstein equations, which we have employed and tested in other studies involving BHs and neutron stars~ (see for instance \cite{Neilsen:2010ax,Neilsen:2014hha}).
We then write down the Klein-Gordon equations 
in terms of the evolution variables of this formulation.

We consider the head-on dynamics of binary, equal-mass, solitonic BSs initially at rest.
We adopt initial data constructed by the superposition of spherically symmetric stars (i.e., as described in the Appendix~\ref{app:BSs}) separated by a distance of $\approx 2.7 R$, implying that the Hamiltonian constraint is only approximately satisfied. 
To extract physical information, we monitor the 
Newman-Penrose $\Psi_4$ radiative scalar, which is computed by contracting the Weyl tensor respectively with a suitably defined null tetrad. This scalar
accounts for the energy carried off by outgoing gravitational waves at infinity.

We adopt finite difference techniques on a regular Cartesian grid to solve the overall system numerically in full 3D without assuming any symmetry. To ensure sufficient resolution in an efficient manner we employ adaptive mesh refinement~(AMR) via the HAD computational infrastructure~\cite{had_webpage} that provides distributed, Berger-Oliger style AMR~\cite{Liebling} with full sub-cycling
in time. A fourth-order accurate spatial discretization satisfying a summation by parts rule, together with a third-order accurate in time Runge-Kutta integration scheme, are used to help
ensure stability of the numerical implementation~\cite{Anderson:2007kz}. We adopt a Courant parameter of $\lambda = 0.25$ so that $\Delta t_l = 0.25 \Delta x_l$ on each refinement level $l$. On each level, one has full sub-cycling in time and therefore ensures that the Courant-Friedrichs-Levy~(CFL) condition dictated by the principal part of
the equations is satisfied. This code has been used extensively for a number of other projects and it has already been rigorously tested. Nevertheless, we have also checked for convergence and Noether charge conservation on the simulations presented in this work.

\bibliographystyle{h-physrev4}
\bibliography{Ref}

\end{document}